\documentclass[%
reprint,
 amsmath,amssymb,
 aps,
]{revtex4-1}

\usepackage{float}
\usepackage{graphicx}% Include figure files
\usepackage{dcolumn}% Align table columns on decimal point
\usepackage{bm}% bold math
\usepackage[flushleft]{threeparttable}
\usepackage{url}
\usepackage{amsmath}
\usepackage{amssymb}
\usepackage{braket}
\usepackage{multirow}
\usepackage{array}
\usepackage{url}
\newcolumntype{P}[1]{>{\centering\arraybackslash}p{#1}}
\usepackage{xcolor}
\usepackage{framed}
\usepackage{lipsum}
\colorlet{shadecolor}{blue!60}

\begin{document}

\preprint{APS/123-QED}

\title{Phase Competition in HfO$_2$ with Applied Electric Field from First Principles}% Force line breaks with \\

\author{Yubo Qi, and Karin M. Rabe}

\affiliation{%
Department of Physics $\&$ Astronomy, Rutgers University, \\
Piscataway, New Jersey 08854, United States
}%

\date{\today}

\begin{abstract}

In this work, the results of first-principles density--functional--theory calculations are used to construct the energy landscapes of HfO$_2$ and its Y and Zr substituted derivatives as a function of symmetry--adapted lattice-mode amplitudes.
These complex energy landscapes possess multiple local minima, corresponding to the tetragonal, o\uppercase\expandafter{\romannumeral3} ($Pca2_1$), and o\uppercase\expandafter{\romannumeral4} ($Pmn2_1$) phases. 
We find that the energy barrier between the non--polar tetragonal phase and the ferroelectric o\uppercase\expandafter{\romannumeral3} phase can be lowered by Y and Zr substitution.
In Hf$_{0.5}$Zr$_{0.5}$O$_2$ with an ordered cation arrangement, Zr substitution makes the o\uppercase\expandafter{\romannumeral4} phase unstable, and it become an intermediate state in the tetragonal to o\uppercase\expandafter{\romannumeral3} phase transition.
Using these energy landscapes, we interpret the structural transformations and hysteresis loops computed for electric-field cycles with various choices of field direction.
The implications of these results for interpreting experimental observations, such as the wake--up and split--up effects, are also discussed.
These results and analysis deepen our understanding of the origin of ferroelectricity and field cycling behaviors in HfO$_2$--based films, and allow us to propose strategies for improving their functional properties.

\end{abstract}

\pacs{Valid PACS appear here}

\maketitle

\section{Introduction}

Hafnia (HfO$_2$) has long been recognized as a high--$\kappa$ material, valuable
in complementary metal--oxide--semiconductor (CMOS) applications~\cite{Bohr07p29,Gutowski02p1897}. 
In 2011, Si--doped HfO$_2$ thin films under mechanically encapsulated crystallization were observed to be ferroelectric~\cite{Boscke11p102903}, 
which led to a renewal of scientific interest and much recent theoretical and experimental research~\cite{Boscke11p102903,Boscke11p112904,Mueller12p123,Mueller12p2412,Muller11p112901,Muller11p114113,Muller12p4318}.

HfO$_2$ is a binary oxide with a number of polymorphs.
At high temperatures ($T>2773$ K), bulk HfO$_2$ has a high--symmetry $Fm\overline{3}m$ cubic fluorite structure,
in which each Hf atom is located at the center of an oxygen cube [Fig.1 (a)].
As the temperature decreases (2773 K$>T>$2073 K)~\cite{Ruh68p23}, this cubic structure becomes unstable.
In this temperature range, the oxygen atoms displace in a $X_2^-$ mode pattern [Fig. S1 (a)] to give a tetragonal $P$4${\rm{_2}}/nmc$ structure [Fig.1 (b)] in which four Hf--O bonds become slightly shorter while the other four become slightly longer~\cite{Reyes14p140103}.
Below 2073 K, bulk HfO$_2$ is found in a $P2_1/c$ monoclinic phase, in which additional distortions reduce the coordination number of Hf from 8 to 7.
However, all these known bulk phases are centrosymmetric and thus nonpolar.
The ferroelectricity observed in HfO$_2$ thin films is attributed to an orthorhombic $Pca$2$_1$ (denoted o\uppercase\expandafter{\romannumeral3}) phase with polarization along [001], 
as has been confirmed by both experimental and theoretical studies~\cite{Lee20p1,Sang15p162905,Park15p1811,Huan14p064111, Reyes14p140103}.
A competing [011]--polarized orthorhombic $Pnm$2$_1$ (denoted o\uppercase\expandafter{\romannumeral4}) phase [Fig.1 (d)] has been proposed on the basis of first--principles calculations ~\cite{Huan14p064111}, but has not yet been experimentally reported.
These facts indicate that HfO$_2$ has a complex energy landscape with multiple local minima.
Investigating the state switching between different local minima, especially switching driven by an electric field, is of great importance in understanding the origin of ferroelectricity and the behavior in applied electric fields.

\begin{figure}[ht]
\includegraphics[width=8.0cm]{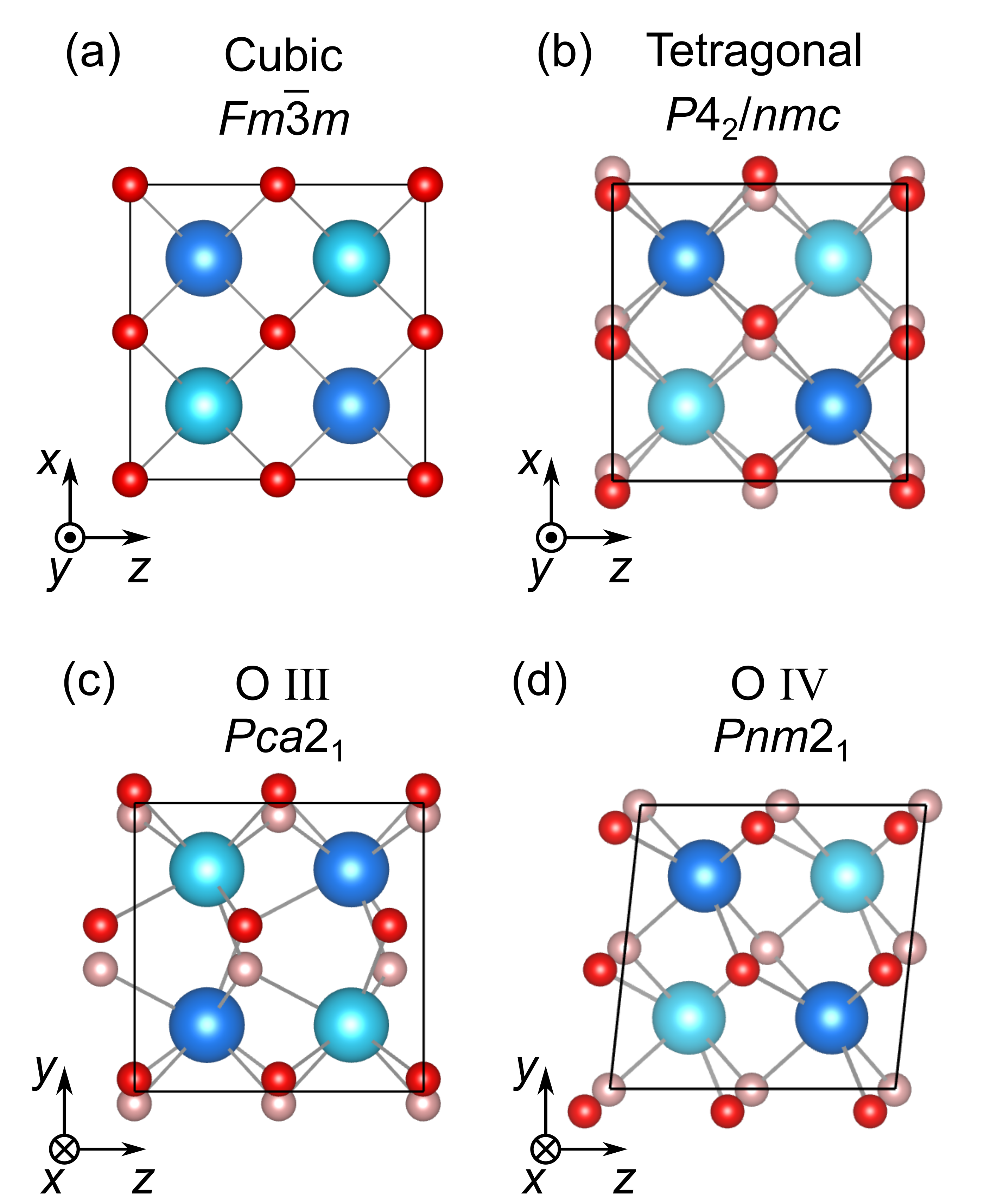}
\caption{(a) Cubic, (b) tetragonal, (c) oIII and (d) oIV phases of HfO$_2$. 
Along the viewing direction, nearer and farther Hf atoms are represented by dark blue and light blue spheres, respectively. 
Similarly, nearer and farther O atoms are represented by red and pink spheres, respectively.}
\end{figure}\label{f1}

In this paper, we report the results of first-principles calculations for HfO$_2$ and substitutional derivatives (Y doped HfO$_2$ and Hf$_{0.5}$Zr$_{0.5}$O$_2$),
with the construction of their energy landscapes as a function of selected symmetry--adapted lattice modes.
Specifically, we consider uniform electric fields applied in various directions with respect to the crystal axes and map out the structural evolution and hysteresis loops for selected electric field profiles.
We find that Y and Zr substitution modifies the energy landscape of pure HfO$_2$.
In particular, the substitutions change the minimum barrier path from the nonpolar tetragonal to the ferroelectric o\uppercase\expandafter{\romannumeral3} phase and between up and down ferroelectric variants.  
These results lead to a better understanding of wake--up and switching behavior in these systems, enabling electric--field control of competing phases in HfO$_2$ and its derivatives.     

\section{Methods}

We carry out Density functional theory (DFT) calculations with the  \textsc{Quantum--espresso}~\cite{Giannozzi09p395502etalp} package for structural relaxations under finite electric fields
and the \textsc{ABINIT}~\cite{Gonze02p478} package for structure optimizations with fixed lattice modes.
Optimized norm--conserving local density approximation (LDA) pseudopotentials were generated using the \textsc{Opium} package~\cite{Opium,Bennett12p14}.
A 4$\times$4$\times4$ Monkhorst--Pack $k$--point mesh was used to sample the Brillouin zone for the conventional 12--atom cell with corresponding meshes for the supercells. The plane--wave cutoff energy was 50 Ry~\cite{Monkhorst76p5188}.
Structural relaxations were performed with a force threshold of 5.0$\times10^{-6}$ Hartree per Bohr.
Relaxed structural parameters for the various bulk phases of pure HfO$_2$ are reported in supplementary materials (SM) section \uppercase\expandafter{\romannumeral1}.
The computed lattice constants are identical to those in our previous work~\cite{Reyes14p140103} and are in good agreement with the results in other studies.

Here, we consider pure HfO$_2$ and two substitutional derivatives: Y doped HfO$_2$ (Y--HfO$_2$) and Hf$_{0.5}$Zr$_{0.5}$O$_2$, as shown in Fig. 2.
The conventional unit cell of pure HfO$_2$ contains 12 atoms or 4 formula units [Fig.2 (a)]. 
For Y--HfO$_2$, we double this unit cell along the $y$ axis and substitute one of the Hf atoms by an Y atom, corresponding to a 12.5~\% doping concentration [Fig. 2 (b)]. 
For Hf$_{0.5}$Zr$_{0.5}$O$_2$, we consider a structure with 1:1 layered cation ordering, as shown in Fig. 2 (c).
Previous experimental work has shown that this superlattice structure with alternating Zr and Hf cation layers 
exhibits electrical behavior quite similar to that of random--cation Hf$_{0.5}$Zr$_{0.5}$O$_2$~\cite{Weeks17p13440}.

\begin{figure}[ht]
\includegraphics[width=8.0cm]{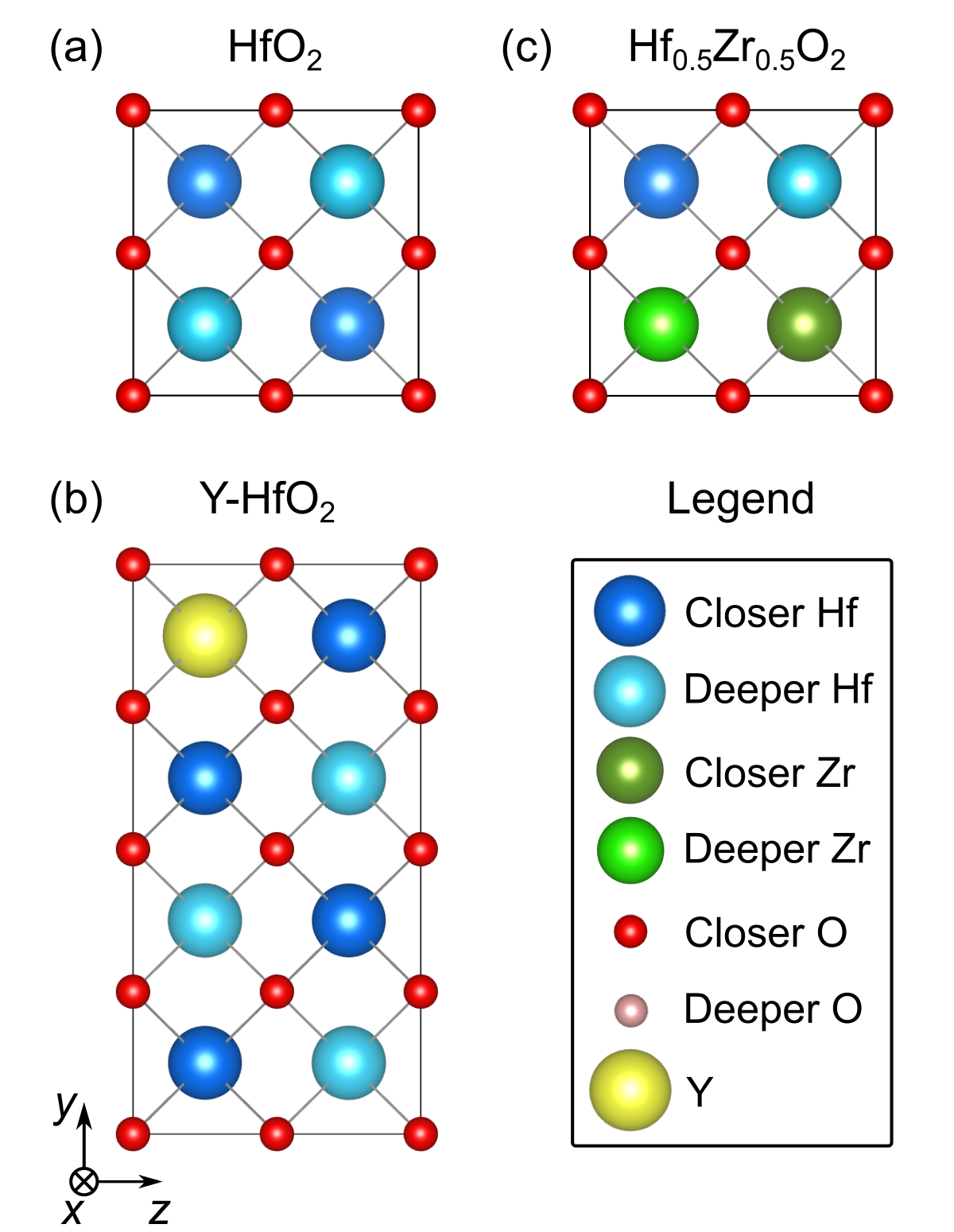}
\caption{High--symmetry cubic structures of the (a) bulk HfO$_2$, (b) Y doped HfO$_2$, and (c) Hf$_{0.5}$Zr$_{0.5}$O$_2$ unit cells.
Hf, Zr, O and Y atoms are represented by blue, green, red and yellow spheres respectively. 
As in Fig. 1, the darkness of the color indicates the spatial distance along the view direction; 
closer atoms are represented by dark colors, which farther ones are represented by light colors.}
\end{figure}\label{f2}

%Mode amplitudes were computed by projecting the atomic displacements in each formula unit onto the mode eigenvectors, which are normalized to 1 and listed in TABLE S6.
We describe the low--symmetry structures in each system by homogeneously straining them back to the lattice of the ideal cubic fluorite reference structure and decomposing the atomic displacements from the reference structure into the symmetry--adapted lattice modes of the reference structure~\cite{Lee20p1,Reyes14p140103}.
The atomic displacement patterns of the lattice modes and the method
for calculating the mode amplitudes are described in SM section \uppercase\expandafter{\romannumeral2}. 
In TABLE S6, the lattice--mode patterns and amplitudes relevant to the optimized tetragonal, o\uppercase\expandafter{\romannumeral3} and o\uppercase\expandafter{\romannumeral4} phases are listed.
The oxygen atom positions in the tetragonal structure are obtained from an anti--polar $X_2^-$ mode, with displacements of oxygen chains along the 4--fold direction, here taken as the $x$ direction, alternating in a 2D checkerboard pattern [Fig. S1(a)].
Starting from the tetragonal structure, the polar phases are generated by the  3--fold--degenerate zone--center polar $\Gamma^-_4$ mode, corresponding to uniform displacement of the Hf atoms relative to the simple cubic oxygen network (Fig. S1(c)). We refer to these modes as  $\Gamma_x$, $\Gamma_y$ and $\Gamma_z$, with amplitudes $u_x$, $u_y$ and $u_z$, respectively. 
The o\uppercase\expandafter{\romannumeral3} structure, whose polarization is along the [001] direction, has only one non--zero polar mode amplitude  $u_z{\neq}\ 0$.
In the o\uppercase\expandafter{\romannumeral4} structure, $u_y=u_z{\neq}\ 0$, indicating a [011]--directed polarization.

The polarization can be estimated from the linear dependence of polarization $P$ on the amplitude of the polar mode in the high--symmetry reference structure as~\cite{Rabe07}
\begin{equation}
P_{i}=\frac{e}{\Omega}Z^*u_{i},
\end{equation}
where $i=x,y,z$, $e$ is the electronic charge, $\Omega$ is the volume of the supercell, and $Z^*$ is the mode Born effective charge (SM section \uppercase\expandafter{\romannumeral3}).

There are two relevant components of the anti--polar $X_{5}^+$ mode. The $X_{5,y}^+ ({\bm{q}}\parallel{\bm{\hat{x}}})$ mode, with amplitude $A_y$, describes the displacement of oxygen layers along the $y$ axis, alternating in the $x$ direction, as shown in Fig. S1 (d).
This amplitude is nonzero both in the o\uppercase\expandafter{\romannumeral3} and o\uppercase\expandafter{\romannumeral4} structures.
The $X_{5,z}^+ ({\bm{q}}\parallel{\bm{\hat{y}}})$ mode, with amplitude $B_z$, describes the displacement of oxygen layers along the $z$ axis, alternating in the $y$ direction, as shown in Fig. S1 (c). This amplitude is nonzero only in the o\uppercase\expandafter{\romannumeral3} structure.
$B_z$ and $u_z$ combine to give zero and non--zero oxygen displacements in alternating layers in the o\uppercase\expandafter{\romannumeral3} structure.
%Later, we will point out that the coupling between $u_z$ and $A_y$ (also $u_z$ and $B_z$) can be viewed as a distinct feature of a specific phase.

For constructing the energy landscape as a function of the lattice--mode amplitudes,
we generate structures with selected amplitudes of the lattice modes, and then, by using the \textsc{ABINIT} package,
relax the structures with the constraint that the selected mode amplitudes stay fixed while the amplitudes of other modes compatible with the resulting symmetry are optimized (SM section \uppercase\expandafter{\romannumeral3}). 
In certain cases to be discussed below, when the lattice mode amplitude is zero, the structure is relaxed with the lower symmetry corresponding to a small nonzero value of the lattice mode amplitude. This ensures that the energy plotted as a function of lattice amplitude is continuous through zero amplitude, which would not in general be the case if the zero-amplitude energy were computed at a saddle point protected by high symmetry.
Then, we fit the calculated energies with Landau polynomials with the lattice--mode amplitudes as order parameters, and use the polynomials to generate the 
energy landscapes.

We determine the effects of uniform applied electric field by relaxing the structure with added forces ${\boldsymbol{F}_i}$ on the ions: 
${\boldsymbol{F}_i}=Z_i^*e\cdot{{\boldsymbol{E}}}$, 
where $Z^*_i$ is the effective charge tensor of ion $i$ and ${\boldsymbol{E}}$ is the applied electric field (SM section \uppercase\expandafter{\romannumeral3}).
To construct electric--field hysteresis loops, an electric field cycle is applied to each of the systems considered with the tetragonal structure as the initial state. 
In each cycle, we fix a direction for the electric field and then increase the magnitude of the electric field in steps from 0 to $+E_{max}$ in steps, then decrease it to $-E_{max}$, and then increase it again to $+E_{max}$. 
Here, $E_{max}=13.5$ MV/cm is the maximum electric field that the HfO$_2$ crystal can sustain in our DFT calculations. Beyond $E_{max}=13.5$, the system becomes ionically conductive and the crystal crashes.
The step size is $\Delta{E}=1.5$ MV/cm, except
near the critical electric field at which switching occurs, where $\Delta{E}$ is reduced to 0.5 MV/cm.
For each value of $E$, the optimized structure from the previous step is used as the starting structure for relaxation.
Hysteresis loops are plotted showing the polar--mode amplitudes of the optimized structures as a function of electric field.

The lattice--mode--amplitude energy landscape can be used to understand the electric--field cycling behaviors to first order in the field. To see this, we start with a given nonzero electric field and relax the structure as described above. 
%If the resulting values of $u_y$ and $u_z$ are constrained and the structure is relaxed at zero field to yield a point in the lattice--mode--amplitude energy landscape, we find that the amplitudes of the nonpolar modes are approximately the same as in the structure computed for the nonzero electric field. 
Next, the resulting structure is relaxed at zero field, constraining the values of $u_y$ and $u_z$, to determine one point in the lattice--mode--amplitude energy landscape. We find that the amplitudes of the nonpolar modes in the structure relaxed under a nonzero electric field, and in the structure computed with a zero electric field and constrained values of $u_y$ and $u_z$, are the same to a good approximation.
Therefore, the effect of nonzero field can be described by adding a term $-P{\cdot}E$, where $P$ is the spontaneous polarization~\cite{Sai02p104108}, to the computed zero-field energy landscape.

\section{Results}

We begin by considering the energy landscape of pure HfO$_2$ in zero applied field, identifying the local minima that correspond to competing phases and analyzing the barriers that separate them.
We then present results for electric--field hysteresis loops of pure HfO$_2$, focusing on the structural changes and switching behavior as the magnitude of the field changes.
Finally, we extend our discussion to the Y--HfO$_2$ and Hf$_{0.5}$Zr$_{0.5}$O$_2$ systems,
and show how Y and Zr substitutions change the energy landscape and electric--field cycling behaviors.

\subsection{Pure HfO$_2$}

The energy landscape of pure HfO$_2$ as a function of the two polar mode amplitudes $u_y$ and $u_z$ is shown in Fig. 3 (a).
There are three types of local minima, corresponding to the tetragonal, o\uppercase\expandafter{\romannumeral3}, and o\uppercase\expandafter{\romannumeral4} phases.
The optimal path from the tetragonal phase to the o\uppercase\expandafter{\romannumeral3} phase runs along the horizontal axis, with the optimal value of $u_y$ at each $u_z$ being zero. Similarly, the optimal path from the tetragonal phase to the o\uppercase\expandafter{\romannumeral4} phase runs along the line $u_y=u_z$. 

\begin{figure}[btbp]
\includegraphics[width=8.0cm]{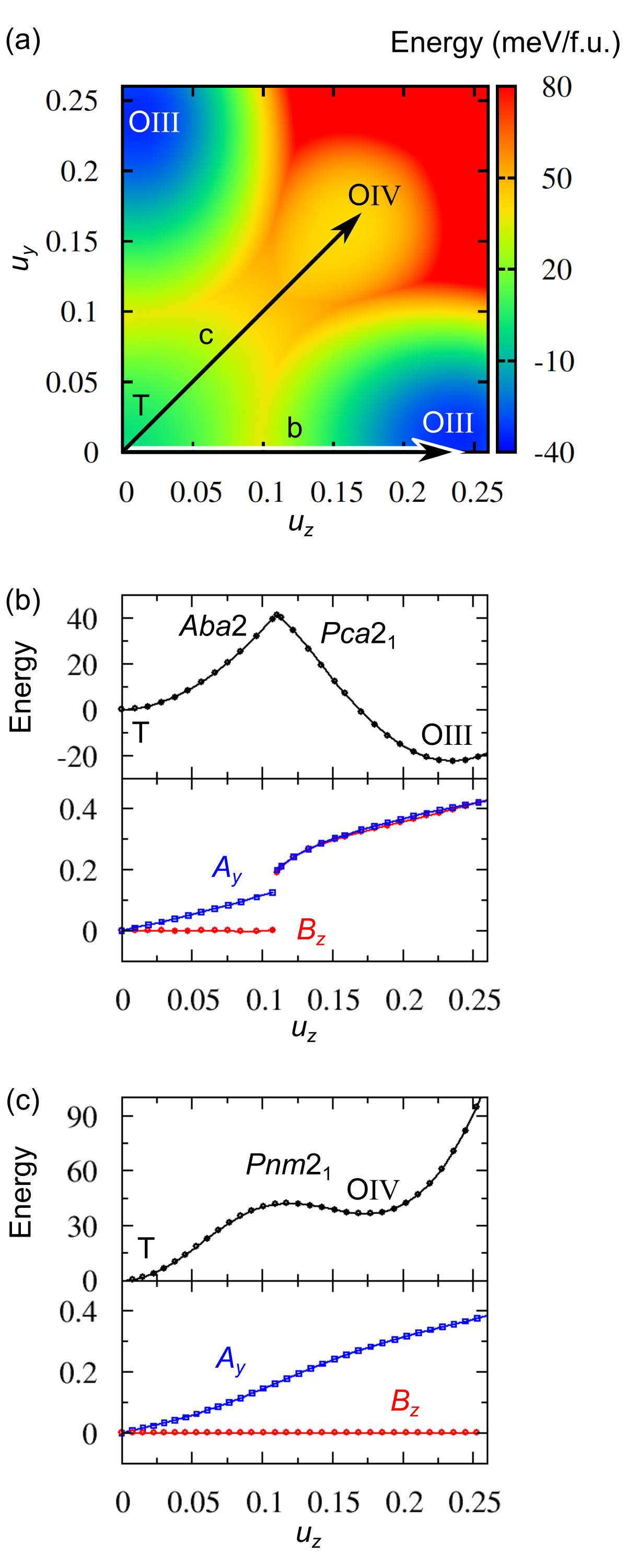}
\caption{(a) Energy landscape of HfO$_2$ as a function of $u_y$ and $u_z$; 
We focus on the two paths indicated by the arrows.  
(b) Energy, $A_y$, and $B_z$ as a function of $u$ along the path b from the tetragonal to the o\uppercase\expandafter{\romannumeral3} phase. Dots are calculated points and the solid line is an interpolation.
(c) Energy, $A_y$, and $B_z$ as a function of $u$ along the path c from the tetragonal to the o\uppercase\expandafter{\romannumeral4} phase.}
\end{figure}\label{f3}

\begin{figure}[htbp]
\includegraphics[width=8.0cm]{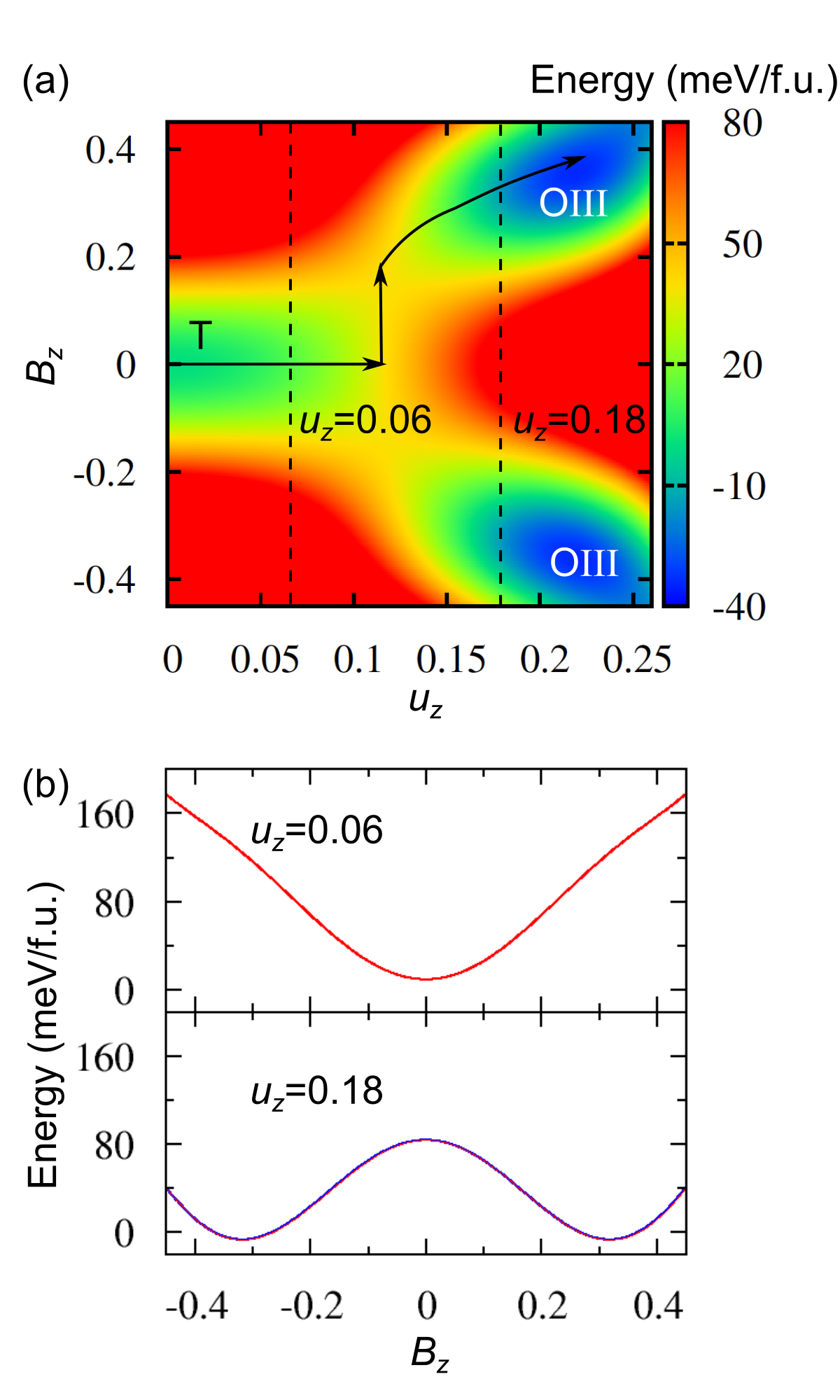}
\caption{(a) Energy landscape of HfO$_2$ as a function of $u_z$ and $B_z$
(b) Energy as a function of $B_z$, for $u_z$ smaller and larger than the critical value 0.12 \AA\  respectively.}
\end{figure}\label{f4}

%tetragonal phase space group is $P$4${\rm{_2}}/nmc$
The nature of the barriers between the local minima can be understood by plotting the energy and the relaxed amplitudes $A_y$ and $B_z$ of the $X_5^+$ modes (Fig. S1) for two straight--line paths in ($u_z$,$u_y$) space, as shown in Fig. 3 (b) and Fig. 3 (c).
Along the straight line path ($u$,0$^+$) [Fig. 3 (b)], a cusp at $u=0.12$ \AA~ separates the two local minima corresponding to the tetragonal and 
o\uppercase\expandafter{\romannumeral3} phases, with a change in space group symmetry from $Aba2$, for the tetragonal phase with a constrained polar distortion, to $Pca2_1$, for the 
o\uppercase\expandafter{\romannumeral3} phase.
In the $Aba2$ phase, the mode amplitude $A_y$ grows linearly with $u$, while $B_z$ remains zero [Fig. 3 (b) and Fig. S1 (b)]. 
In the $Pca2_1$ phase, found for $u> 0.12$ \AA, $A_y\ \approx \ B_z\neq0$, with jumps in both $A_y$ and $B_z$ across the cusp.
The jump in $B_z$ can be understood by plotting the energy landscape as a function of $u_z$ and $B_z$, with other symmetry--allowed modes, including $A_y$, allowed to relax [Fig. 4 (a)]. 
For $u_z<0.12$ \AA, the energy as a function of $B_z$ has a single minimum at $B_z$=0, while for $u_z>0.12$ \AA, it is a double well [Fig. 4 (b)].

To understand the nature of the barrier between the tetragonal phase and the o\uppercase\expandafter{\romannumeral4} phase, we show the energy and the mode amplitudes $A_y$ and $B_z$ as a function of $u$ along the straight--line path ($u,u$) connecting the tetragonal phase to the o\uppercase\expandafter{\romannumeral4} phase [Fig. 3 (b)]. 
Nonzero $u$ lowers the symmetry to $Pnm2_1$. With increasing $u$, $A_y$ grows linearly with $u_z$ and $B_z$ remains zero. The $Pnm2_1$ o\uppercase\expandafter{\romannumeral4} phase is metastable, with an energy of 31 meV/f.u. above the tetragonal phase.  The energy barrier from the  o\uppercase\expandafter{\romannumeral4} phase is only 6 meV/f.u. and is smooth rather than cusp--like.

To investigate the electrical field cycling behavior, we applied two electric field cycles to the tetragonal phase, one with electric field along the [001] direction and the other along [011]. 
The polar--mode--amplitude hysteresis loops are shown in Fig. 5 (a) and the evolution of the free energy landscape with electric field is shown in Figs. S4 and S5. 
For $\bm{E}//$[001], a tetragonal to polar o\uppercase\expandafter{\romannumeral3} phase transition occurs at 
$E_{\rm{T}{\rightarrow}{\rm{P}}}=11$ MV/cm. 
After the electric field is returned to zero, the system is trapped at the local minimum corresponding to the o\uppercase\expandafter{\romannumeral3} structure.
As the electric field becomes more negative, the polarization switches at $E_{\rm{P}{\rightarrow}-{\rm{P}}}=-13$ MV/cm.
Note that after the initial transition to the o\uppercase\expandafter{\romannumeral3} phase, field cycling switches the system back and forth between up and down variants as in a conventional ferroelectric.
For $\bm{E}//$[011], 
the critical field for inducing a tetragonal to polar o\uppercase\expandafter{\romannumeral4} phase transition is $E_{\rm{T}{\rightarrow}{\rm{P}}}=7$ MV/cm.
After the electric field is returned to zero, the system is trapped at the local minimum corresponding to the o\uppercase\expandafter{\romannumeral4} structure.
As the electric field becomes more negative, this local minimum disappears and at $E_{\rm{P}{\rightarrow}{\rm{T}}}=-2.5$ MV/cm, the system switches back to the tetragonal phase. With further increase of the electric field in the negative direction, the polarization switches at $E_{\rm{P}{\rightarrow}-{\rm{P}}}=-7$ MV/cm.
%The resulting double hysteresis loop resembles that of a conventional antiferroelectric.

 \begin{figure*}
 \centering
\includegraphics[width=18.0cm]{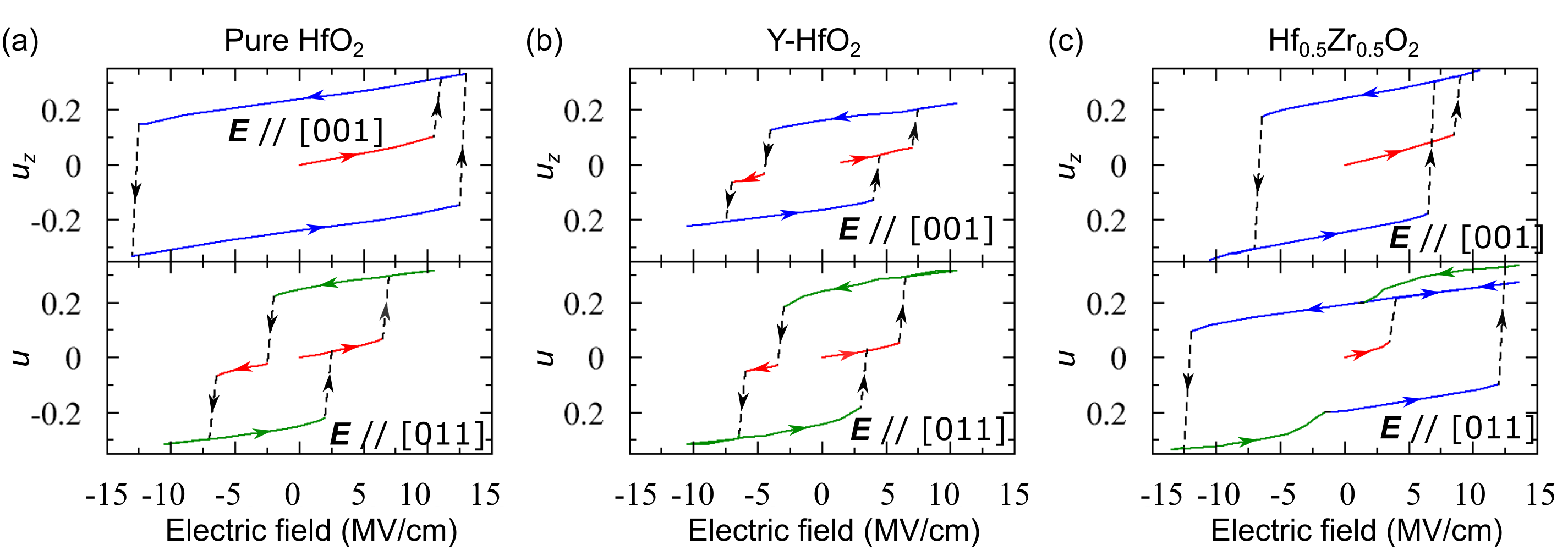}
\caption{Hysteresis loops of (a) pure HfO$_2$, (b) Y--HfO$_2$ and (c) Hf$_{0.5}$Zr$_{0.5}$O$_2$, with electric field along [001] and [011] directions respectively.
The evolution of the free energy landscape with electric field is shown in SM section \uppercase\expandafter{\romannumeral5}.
In pure HfO$_2$, the critical field for polarization flipping ($E_{\rm{P}{\rightarrow}-{\rm{P}}}=-13$ MV/cm) is very close to the maximum electric field which the crystal can sustain ($E_{max}=13.5$ MV/cm).
In the cases with ${\rm{E}}//[011]$, $u$ represents the magnitude of the polar modes projected on the direction of the electric field. 
The color of the segment (red, blue and green) indicates the state is in the local minimum corresponding to the tetragonal, o\uppercase\expandafter{\romannumeral3}, o\uppercase\expandafter{\romannumeral4}
structure respectively. For Y--HfO$_2$, the blue segment represents that the structure containing an o\uppercase\expandafter{\romannumeral3} part.
}
\end{figure*}\label{f5}

%Hysteresis loops for Y--HfO$_2$ and Hf$_{0.5}$Zr$_{0.5}$O$_2$ are also plotted [Fig. 5 (b) and (c)]. 
%In the following part, we will illustrate that their differences can be explained by the change of energy landscape upon doping.

\subsection{Y--HfO$_2$}

The Y--doped HfO$_2$ 1$\times$2$\times$1 supercell is shown in Fig. 2 (b).
The arrangement of Y atoms considered respects tetragonal symmetry, so that relaxing the structure from the high--symmetric cubic phase gives rise to a nonpolar tetragonal structure quite similar to the tetragonal structure of HfO$_2$, with nonzero values for the $X_2^-$ mode. Here, we would like to note that for the phases we study, we find that heterovalent Y doping adds a hole at the top of he valence band without introducing any states in the gap, which is consistent with Ref.~\cite{Lee08p012102}.

The energy of Y--HfO$_2$ as a function of $u_z$ is shown in Fig. 6 (a), where we see three phases separated by cusps, with one local minimum at $u_z=0$ \AA\ and a second at $u_z=0.17$ \AA.
Relative to pure HfO$_2$, both the value of $u_z$ at the first cusp and the height of the energy barrier separating the two local minima are lower in Y--HfO$_2$. 
For values of $u_z$ below the first cusp value $u_z<0.06$ \AA, the energy as a function of $u_z$ is very close to that of pure HfO$_2$. In both systems,
with increasing $u_z$, $A_y$ couples with $u_z$ linearly and $B_z$ remains zero.
%For values of $u_z$ above the cusp value, the energy profile of Y--HfO$_2$ can be divided into two segments.
At the first cusp $u_z=0.06$ \AA, both $A_y$ and $B_z$ exhibit boosts, indicating a structural transformation with a change in symmetry.
A representative structure with $u_z=0.08$ \AA\ is shown in Fig. 6 (b).
The supercell can be divided into two parts with distinct behavior.
The unsubstituted part of the supercell has the atomic displacements characteristic of the polar o\uppercase\expandafter{\romannumeral3} structure, which is responsible for the ferroelectricity.
Here, the adjacent oxygen layers (indicated by the arrows) have different bonding environments.
As a result, they are prone to have different displacements, leading to a non--zero $B_z$, which is the character of the o\uppercase\expandafter{\romannumeral3} structure.
The part of the supercell containing the Y atom is less distorted.
At the second cusp $u_z=0.10$ \AA, there is a downward jump in $B_z$, indicating another phase transition.
%, but without a symmetry change.
Fig. 6 (b) shows the structure of the polar phase at $u=u_{min}=0.18$ \AA. The structure in the unsubstituted part of the supercell closely resembles the polar o\uppercase\expandafter{\romannumeral3} structure, while the part of the supercell containing the Y atom is close to the o\uppercase\expandafter{\romannumeral4} phase.
The decrease in $B_z$ is attributed to the formation of the o\uppercase\expandafter{\romannumeral4} phase in the Y-substituted part, with zero $B_z$ as indicated in Fig. 3 (c).
 
Relative to pure HfO$_2$, both the value of $u_z$ at the cusp and the height of the energy barrier in Y--HfO$_2$ are lower. indicating a smaller critical electric field $E_{\rm{T}{\rightarrow}{\rm{P}}}$ for triggering the tetragonal to polar phase transition.
The behavior in applied electric field exhibits corresponding differences. The detailed evolution of the free energy landscape with electric field is shown in Fig. S6. Here, because of the added hole, the density of states at the Fermi level is nonzero and we investigate the change of structure under an electric field with an accompanying nonzero field--induced current.
As shown in Fig. 5 (b), for ${\rm{E}}//[001]$, the critical field $E_{\rm{T}{\rightarrow}{\rm{P}}}=7.5$ MV/cm for the transition from the tetragonal to the polar phase is much reduced
compared with that [$E_{\rm{T}{\rightarrow}{\rm{P}}}=11$ MV/cm] in pure HfO$_2$, as expected from the lower barrier.
Futhermore, the saturation polarization is smaller than in pure HfO$_2$.

Fig. 7 shows the energy profile of Y--HfO$_2$ as a function of $u=u_y=u_z$. For small $u$, $A_y$ couples with $u_z$ linearly while $B_z$ remains approximately zero, very close to the behavior in pure HfO$_2$.
Similar to the energy profile of Y--HfO$_2$ along the [001] direction, Y doping also introduces intermediate states, here two rather than one, and the phase changes along the transformation path are indicated by the singular points of the $B_z$ profile (see SM Fig. S3 for the representative structure of each intermediate state).
Even though the relative energy of the o\uppercase\expandafter{\romannumeral4} structure is lowered in Y--HfO$_2$,
we see that the curvatures of the local minima, which correlate to the maximum slopes in the energy profile, are approximately the same as those in pure HfO$_2$, indicating that the critical fields for inducing the phase transitions are similar [Fig. 5 (a) and (b)].

\begin{figure}[htbp]
\includegraphics[width=8.0cm]{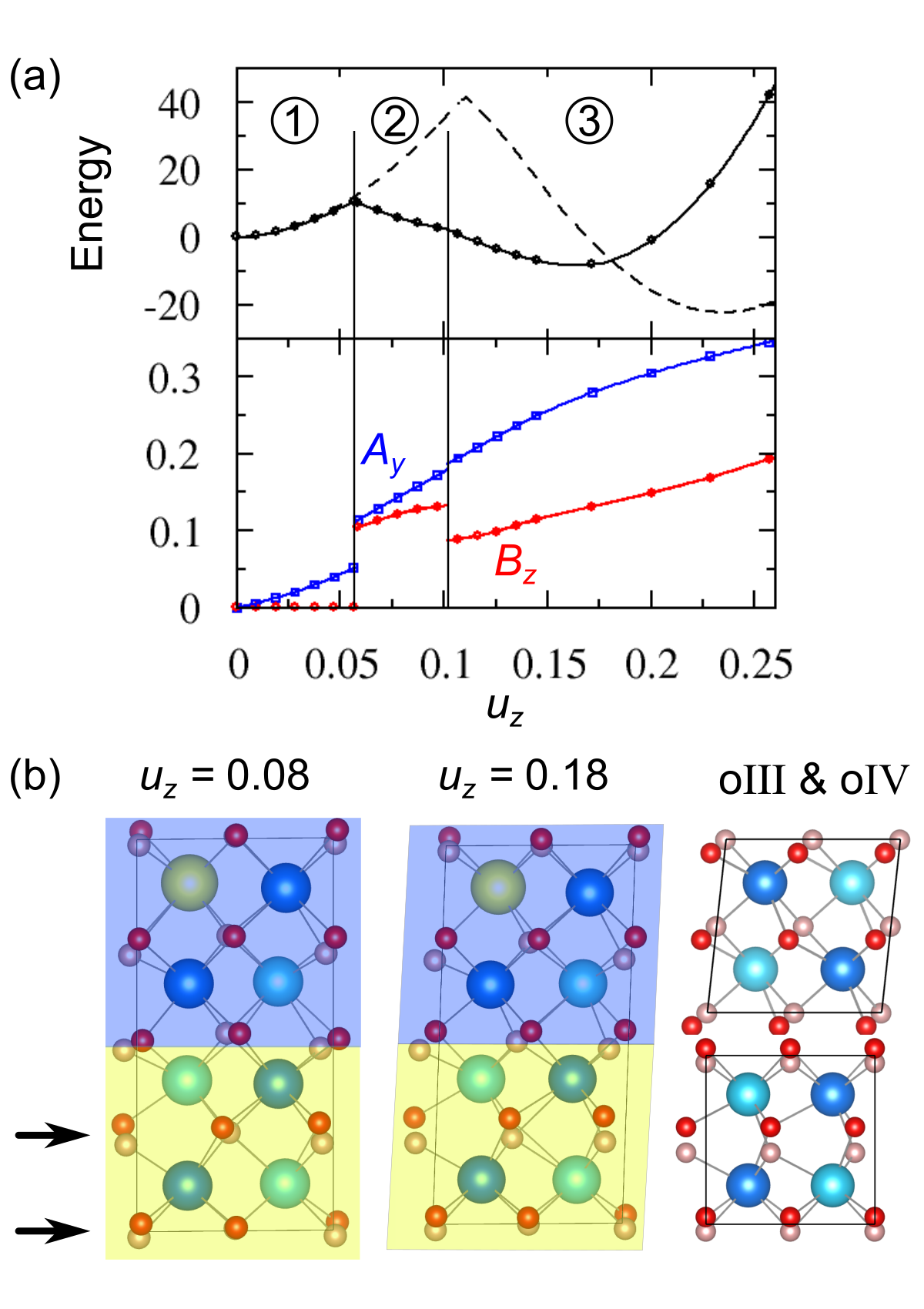}
\caption{(a) Energy of Y--HfO$_2$ as a function of $u_z$ with other degrees of freedom relaxed. The same profile for pure HfO$_2$ is also plotted (dashed line) for comparison.
%For each compound, structures with different $u_z$ are constructed based on their tetragonal structures, by artificially enlarging the cation--displacements. 
%Then structural optimizations are carried out with fixed $u_z$.
(b) Crystal structure of Y--HfO$_2$ for $u_z$=0.08 and 0.18 \AA. The structures of the o\uppercase\expandafter{\romannumeral3} and o\uppercase\expandafter{\romannumeral4} phase in pure HfO$_2$ are also shown for comparison.}
\end{figure}\label{f6}

\begin{figure}[htbp]
\includegraphics[width=8.0cm]{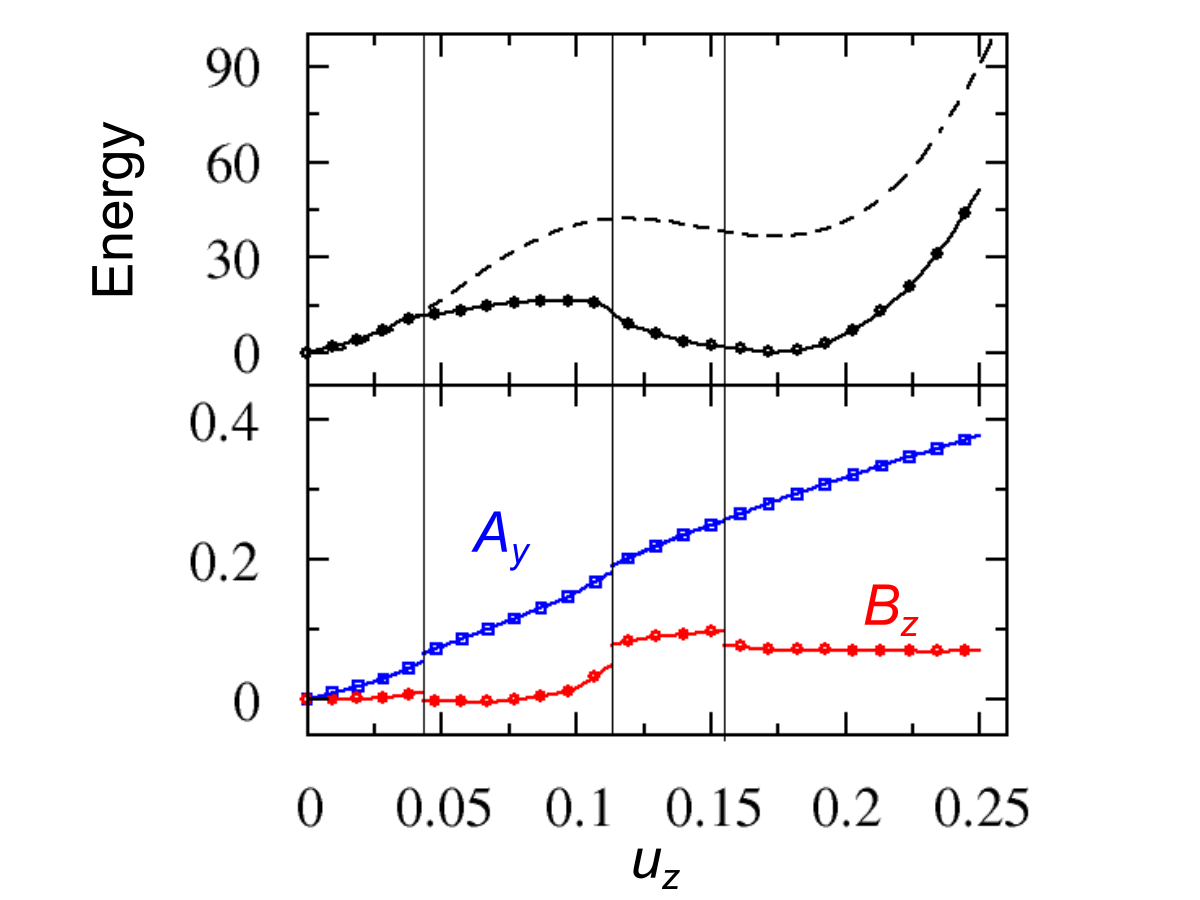}
\caption{Energy, $A_y$, and $B_z$ as a function of $u_z$ along the path of the tetragonal to o\uppercase\expandafter{\romannumeral4} phase transition.
Along this transformation path, $u_y$ is fixed to be equal to $u_z$. The representative structures of each intermediate states are shown in Fig. S3.}
\end{figure}\label{f7}

\subsection{Hf$_{0.5}$Zr$_{0.5}$O$_2$}

The energy landscape of Hf$_{0.5}$Zr$_{0.5}$O$_2$ as a function of the two polar mode amplitudes $u_y$ and $u_z$ is shown in Fig. 8 (a).
As in pure HfO$_2$, there are local minima corresponding to the tetragonal and o\uppercase\expandafter{\romannumeral3} phases.
However, the cation layering normal to the $y$ direction breaks the $y$--$z$ symmetry of pure HfO$_2$, so the two local minima for o\uppercase\expandafter{\romannumeral3}, with polarization in the y direction and polarization in the $z$ direction respectively, are not equivalent.
In addition, the polarization of z--polarized o\uppercase\expandafter{\romannumeral3} polarized in the $z$ direction has a small nonzero $y$ component, which resulting in two symmetry--related variants, one with positive $y$ component and one with negative $y$ component.
We also see that in Hf$_{0.5}$Zr$_{0.5}$O$_2$, 
the o\uppercase\expandafter{\romannumeral4} phase is unstable.
Structural optimization starting from an o\uppercase\expandafter{\romannumeral4} structure leads to the o\uppercase\expandafter{\romannumeral3} minimum, as shown in Fig. 8 (a).

In Fig. 8 (b), we plot 
the energy of Hf$_{0.5}$Zr$_{0.5}$O$_2$ as a function of $u_z$, with the amplitudes of other modes at their optimized values, shown in Fig. 8 (a) and (b). 
As in pure HfO$_2$, this energy profile also exhibits two local minima corresponding to the tetragonal and o\uppercase\expandafter{\romannumeral3} phases, very little changed from the pure case.
However, the cusp seen in pure HfO$_2$ is cut off, substantially lowering the energy barrier between the two phases.
The lowering of the energy barrier results from the change in the path through the energy landscape, which goes through the o\uppercase\expandafter{\romannumeral4} phase region, and around the peak between the tetragonal and o\uppercase\expandafter{\romannumeral3} phase [see Fig. 8 (a)]. This is in contrast to pure HfO$_2$, in which the optimal path runs along the horizontal axis.

These features significantly influence the electric--field cycling behavior of Hf$_{0.5}$Zr$_{0.5}$O$_2$.
As shown in Fig. 5 (c), both [001] and [011] directed electric fields induce a transition from the tetragonal phase to the $z$--polarized  o\uppercase\expandafter{\romannumeral3} phase,
with the critical field $E_{\rm{T}{\rightarrow}{\rm{P}}}$ being much smaller if $E$ is applied along the [011] direction (3.5 MV/cm {\em{vs.}} 9 MV/cm).
Furthermore, the hysteresis loop for $E//[011]$ exhibits a complex and unconventional shape.
Once an [011]--directed electric field has induced the tetragonal to $z$--polarized o\uppercase\expandafter{\romannumeral3} phase transition, the system stays trapped in the local  o\uppercase\expandafter{\romannumeral3} minimum as the field is increased to $E=13.5$ MV/cm and then decreased. 
However, when the electric field in the reverse direction reaches $E=-12.5$ MV/cm, the polarization switches to the [011] direction to align with the field driving the system into the o\uppercase\expandafter{\romannumeral4} phase polarized in the [0$\bar{1}\bar{1}$] direction.
As the magnitude of the field is decreased, the system remains trapped in the local o\uppercase\expandafter{\romannumeral4} minimum until the field reaches $-1$ MV/cm, at which point the system switches back to the o\uppercase\expandafter{\romannumeral3} phase with polarization in the negative $z$--direction.  In Fig. S8, we show this in more detail, connecting the complexity in the hysteresis loop of Hf$_{0.5}$Zr$_{0.5}$O$_2$ to the specific features of its energy landscape.

The surprising fact that the critical field for switching to the o\uppercase\expandafter{\romannumeral3} phase is lower along [011] than along the polarization direction [001] is related to the complexity of the energy landscape.
%Here, we would like to propose that the coercive field also depends on the direction of the electric field driving the state out of the o\uppercase\expandafter{\romannumeral3} minimum. 
Indeed, an electric field along [01$\bar{1}$] is also more effective in driving the state out of the o\uppercase\expandafter{\romannumeral3} minimum, compared with a [00$\bar{1}$] or [0$\bar{1}\bar{1}$] oriented one.  
In Fig 8(c), we show the polar mode amplitude hysteresis loop for an electric field cycle in which the field is first increased from zero along [011], decreased to zero, and then increased along [01$\bar{1}$]. Because the energy landscape is the same under $u_z{\to}-u_z$, the hysteresis loop is symmetrical.
The magnitude of the coercive field for flipping the polarization is 6.5 MV/cm, smaller than those (7 and 12.5 MV/cm) with electric field along [001] and [011] directions.
These results open possibilities for improving the electrical performance of HfO$_2$ by modulating the direction of applied electric field. 

We note that the results we have presented are based on calculations with the LDA functional. Using the generalized gradient approximation (GGA) changes the relative energy of each phase (see SM section \uppercase\expandafter{\romannumeral1}, TABLE S4), but does not destabilize any of the phases we considered. Therefore, the energy landscapes generated by GGA possess the same multiple local minima, with an electric field similarly inducing phase transitions between them. 

\section{Discussion}

We have demonstrated that the tetragonal phase of HfO$_2$ can transform to the ferroelectric o\uppercase\expandafter{\romannumeral3} or o\uppercase\expandafter{\romannumeral4} phase under an electric field, providing insights into the emergence of ferroelectricity and electric-field-cycling behaviors.
In the following subsections, we will relate our computational results to experimental observations.

\subsection{Wake--up and ferroelectricity}

\begin{figure}[H]
\includegraphics[width=8.0cm]{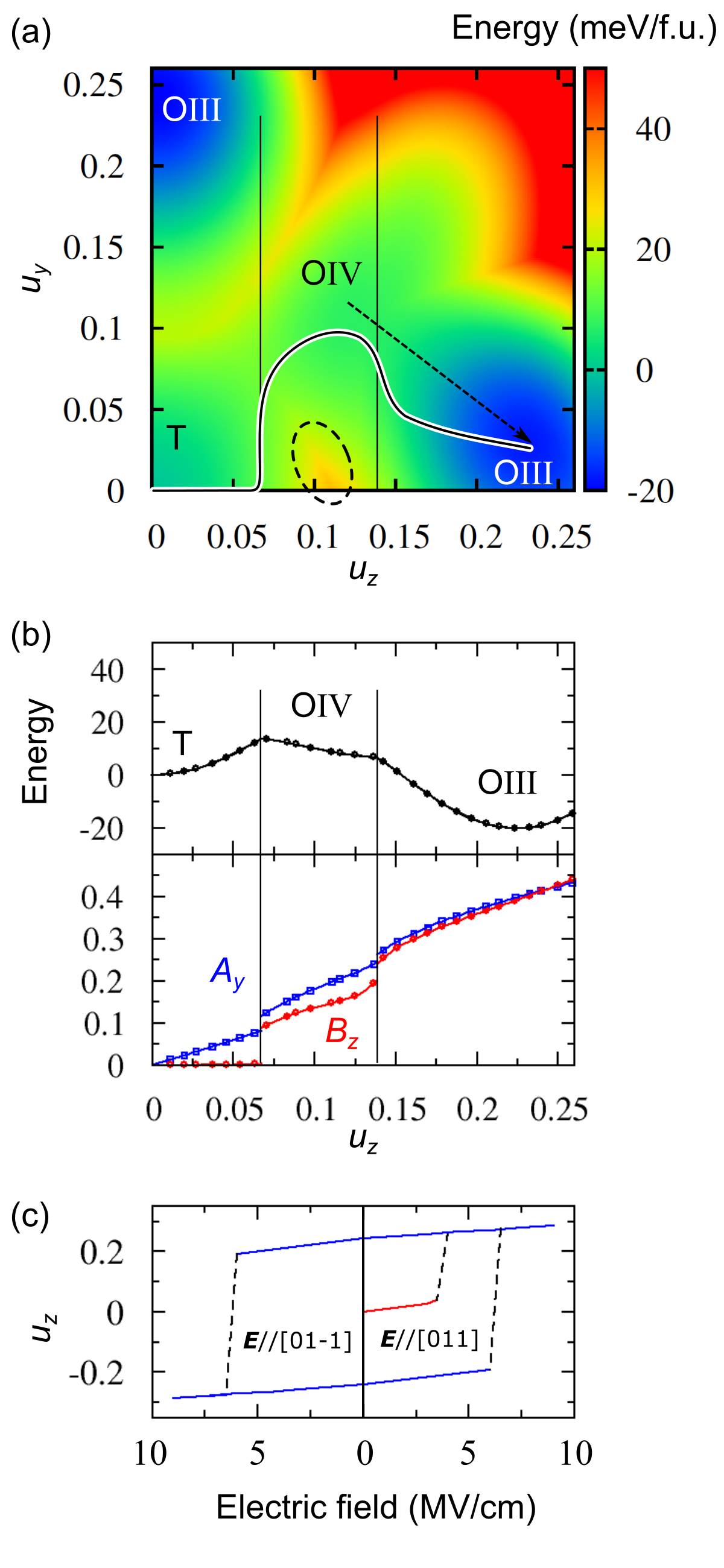}
\caption{(a) Energy landscape of Hf$_{0.5}$Zr$_{0.5}$O$_2$ with respect to the amplitudes of the $u_y$ and $u_z$ polar modes;
(b) Energy profiles  of Hf$_{0.5}$Zr$_{0.5}$O$_2$ as a function of $u_z$，with energy minimized with respect to $u_y$. 
The values of $u_y$ at each $u_z$ are shown by the black curve with a white edge in (a). The values of $A_y$ and $B_z$ at each $u_z$ are shown in the lower panel.
(c) Hysteresis loop in which the forward electric is along the [011] direction, and the backward one is along [01-1].}
\end{figure}\label{f8}

The wake--up effect, which refers to the fact that an as--grown HfO$_2$--based thin film needs several electric field cycles to establish its full value of switching polarization, has been widely observed in experiments~\cite{Zhou2013p192904,Pevsic16p4601,Grimley16p1600173,Martin14p8198}.
This wake--up effect has been attributed to various factors, including structural change at the interface~\cite{Hoffmann15p072006,Grimley16p1600173,Pevsic16p4601}, and the internal bias fields caused by oxygen vacancies ~\cite{Schenk15p20224,Schenk14p19744,Mart19p2612}.

In addition, an electric--field--induced nonpolar to polar phase transition is expected to play a significant role~\cite{Pevsic16p4601,Hoffmann15p154,Park15p192907,Batra17p4139}.
In experiments, doped HfO$_2$ films are usually deposited at high temperature~\cite{Boscke11p102903,Boscke11p112904,Mueller12p123,Muller12p4318}. High temperature, surface energy and dopants all promote the formation of the tetragonal phase~\cite{Batra16p172902,Lee08p012102}.
Upon cooling, the thin films can be kinetically trapped in the tetragonal phase, since it is metastable, with no unstable phonon modes~\cite{Reyes14p140103}. However, the o\uppercase\expandafter{\romannumeral3} structure has a lower energy than the nonpolar tetragonal phase, 
so that once the more stable o\uppercase\expandafter{\romannumeral3} structure is induced by an electric field, a electric field opposite to the polarization will tend to flip the polarization to the corresponding o\uppercase\expandafter{\romannumeral3} phase
rather than drive the system back to the tetragonal phase.

In our calculations,
for pure HfO$_2$ the critical electric field $E_{\rm{T}{\rightarrow}{\rm{P}}}$ for triggering this phase transition is 11 MV/cm.
Our calculations also show that Y and Zr doping can cut off the cusp in the energy profiles by introducing intermediate states, and thus lower the energy barrier and critical fields ($E_{\rm{T}{\rightarrow}{\rm{P}}}=7.5$ MV/cm in Y--HfO$_2$ and 4 MV/cm in Hf$_{0.5}$Zr$_{0.5}$O$_2$).
Moreover, we note that since the energy barriers are not very high, 
thermal vibration can facilitate the phase transition and further lower the critical fields.
With the simple assumption that a phase transition will be driven by thermal vibration if (1) compared with the current state, the energy of another state is lower by more than $k_BT/2$ per formula unit, 
and (2) the energy barrier between the two states is lower than $k_BT/2$ per formula unit, we find that the critical fields are further reduced to 1.1 MV/cm (in Y--HfO$_2$) and 2.2 MV/cm (in Hf$_{0.5}$Zr$_{0.5}$O$_2$),
which are the same order of magnitude as the experimental values~\cite{Boscke11p112904,Boscke11p102903,Muller12p4318}.

\subsection{Multiple jumps in the $P$--$E$ loop}

In a conventional ferroelectric hysteresis loop, the polarization jumps at the coercive field.
However, in HfO$_2$ based thin films, jumps in polarization can occur at two or more values of the electric field, leading to an irregular shape of the $P$--$E$ loop (as in Fig. 5).
This phenomenon has been discussed in Ref.~\cite{Schenk14p19744}, and referred as the `split--up' effect. 
This `split--up' effect has been attributed to the different activation energies in grains with different oxygen vacancy concentrations~\cite{Schenk14p19744}.
Here, we would like to point out that the competing phases in the energy landscape can also contribute to this effect.
As shown in Fig. 5, state hopping between different (tetragonal, o\uppercase\expandafter{\romannumeral3} and o\uppercase\expandafter{\romannumeral4}) local minima can lead to multiple boosts.
Moreover, our results demonstrate that electric field cycling behaviors are also field--direction dependent. In HfO$_2$-based thin films composed of multiple grains~\cite{Grimley16p1600173,Hoffmann16p8643,Martin14p8198}, grains with different lattice orientations may follow different transformation paths with different critical fields.

\section{Conclusion}

In this study, we investigated the energy landscapes of pure HfO$_2$, Y--doped HfO$_2$, and Hf$_{0.5}$Zr$_{0.5}$O$_2$ as a function of the polar modes $u_y$ and $u_z$. The complex energy landscapes of these systems are found to possess multiple local minima, corresponding to the tetragonal, o\uppercase\expandafter{\romannumeral3}, and o\uppercase\expandafter{\romannumeral4} phases respectively. Moreover,
we find that Y and Zr doping can lower the energy barriers between the non--polar tetragonal phase and the ferroelectric o\uppercase\expandafter{\romannumeral3} phase, by introducing intermediate states. 
In Hf$_{0.5}$Zr$_{0.5}$O$_2$ with an ordered cation arrangement, the o\uppercase\expandafter{\romannumeral4} phase becomes unstable, and
a new transformation path for the tetragonal to o\uppercase\expandafter{\romannumeral3} phase transition is opened, with a lower energy barrier.
We connect our results to experimental observations, such as the wake--up effect and $P$--$E$ loop with irregular shape, and also propose dependence of behavior on the direction of applied electric field.
This work provides useful insights about the origin of the ferroelectric phase in HfO$_2$--based films, and suggests strategies for improving their ferroelectric performance.

\section*{ACKNOWLEDGMENTS}
We thank Sebastian E. Reyes--Lillo, Fei--Ting Huang, Xianghan Xu, and Sang--Wook Cheong for valuable discussions.
This work was supported by Office of Naval Research Grant N00014--17--1--2770.
Computations were performed by using the resources provided by the High--Performance Computing Modernization Office of the Department of Defense and the Rutgers University Parallel Computing (RUPC) clusters.

\bibliography{cites}

\begin{thebibliography}{37}%
\makeatletter
\providecommand \@ifxundefined [1]{%
 \@ifx{#1\undefined}
}%
\providecommand \@ifnum [1]{%
 \ifnum #1\expandafter \@firstoftwo
 \else \expandafter \@secondoftwo
 \fi
}%
\providecommand \@ifx [1]{%
 \ifx #1\expandafter \@firstoftwo
 \else \expandafter \@secondoftwo
 \fi
}%
\providecommand \natexlab [1]{#1}%
\providecommand \enquote  [1]{``#1''}%
\providecommand \bibnamefont  [1]{#1}%
\providecommand \bibfnamefont [1]{#1}%
\providecommand \citenamefont [1]{#1}%
\providecommand \href@noop [0]{\@secondoftwo}%
\providecommand \href [0]{\begingroup \@sanitize@url \@href}%
\providecommand \@href[1]{\@@startlink{#1}\@@href}%
\providecommand \@@href[1]{\endgroup#1\@@endlink}%
\providecommand \@sanitize@url [0]{\catcode `\\12\catcode `\$12\catcode
  `\&12\catcode `\#12\catcode `\^12\catcode `\_12\catcode `\%12\relax}%
\providecommand \@@startlink[1]{}%
\providecommand \@@endlink[0]{}%
\providecommand \url  [0]{\begingroup\@sanitize@url \@url }%
\providecommand \@url [1]{\endgroup\@href {#1}{\urlprefix }}%
\providecommand \urlprefix  [0]{URL }%
\providecommand \Eprint [0]{\href }%
\providecommand \doibase [0]{http://dx.doi.org/}%
\providecommand \selectlanguage [0]{\@gobble}%
\providecommand \bibinfo  [0]{\@secondoftwo}%
\providecommand \bibfield  [0]{\@secondoftwo}%
\providecommand \translation [1]{[#1]}%
\providecommand \BibitemOpen [0]{}%
\providecommand \bibitemStop [0]{}%
\providecommand \bibitemNoStop [0]{.\EOS\space}%
\providecommand \EOS [0]{\spacefactor3000\relax}%
\providecommand \BibitemShut  [1]{\csname bibitem#1\endcsname}%
\let\auto@bib@innerbib\@empty
%</preamble>
\bibitem [{\citenamefont {Bohr}\ \emph {et~al.}(2007)\citenamefont {Bohr},
  \citenamefont {Chau}, \citenamefont {Ghani},\ and\ \citenamefont
  {Mistry}}]{Bohr07p29}%
  \BibitemOpen
  \bibfield  {author} {\bibinfo {author} {\bibfnamefont {M.~T.}\ \bibnamefont
  {Bohr}}, \bibinfo {author} {\bibfnamefont {R.~S.}\ \bibnamefont {Chau}},
  \bibinfo {author} {\bibfnamefont {T.}~\bibnamefont {Ghani}}, \ and\ \bibinfo
  {author} {\bibfnamefont {K.}~\bibnamefont {Mistry}},\ }\href@noop {}
  {\bibfield  {journal} {\bibinfo  {journal} {IEEE spectrum}\ }\textbf
  {\bibinfo {volume} {44}},\ \bibinfo {pages} {29} (\bibinfo {year}
  {2007})}\BibitemShut {NoStop}%
\bibitem [{\citenamefont {Gutowski}\ \emph {et~al.}(2002)\citenamefont
  {Gutowski}, \citenamefont {Jaffe}, \citenamefont {Liu}, \citenamefont
  {Stoker}, \citenamefont {Hegde}, \citenamefont {Rai},\ and\ \citenamefont
  {Tobin}}]{Gutowski02p1897}%
  \BibitemOpen
  \bibfield  {author} {\bibinfo {author} {\bibfnamefont {M.}~\bibnamefont
  {Gutowski}}, \bibinfo {author} {\bibfnamefont {J.~E.}\ \bibnamefont {Jaffe}},
  \bibinfo {author} {\bibfnamefont {C.-L.}\ \bibnamefont {Liu}}, \bibinfo
  {author} {\bibfnamefont {M.}~\bibnamefont {Stoker}}, \bibinfo {author}
  {\bibfnamefont {R.~I.}\ \bibnamefont {Hegde}}, \bibinfo {author}
  {\bibfnamefont {R.~S.}\ \bibnamefont {Rai}}, \ and\ \bibinfo {author}
  {\bibfnamefont {P.~J.}\ \bibnamefont {Tobin}},\ }\href@noop {} {\bibfield
  {journal} {\bibinfo  {journal} {Appl. Phys. Lett.}\ }\textbf {\bibinfo
  {volume} {80}},\ \bibinfo {pages} {1897} (\bibinfo {year}
  {2002})}\BibitemShut {NoStop}%
\bibitem [{\citenamefont {B{\"o}scke}\ \emph
  {et~al.}(2011{\natexlab{a}})\citenamefont {B{\"o}scke}, \citenamefont
  {M{\"u}ller}, \citenamefont {Br{\"a}uhaus}, \citenamefont {Schr{\"o}der},\
  and\ \citenamefont {B{\"o}ttger}}]{Boscke11p102903}%
  \BibitemOpen
  \bibfield  {author} {\bibinfo {author} {\bibfnamefont {T.}~\bibnamefont
  {B{\"o}scke}}, \bibinfo {author} {\bibfnamefont {J.}~\bibnamefont
  {M{\"u}ller}}, \bibinfo {author} {\bibfnamefont {D.}~\bibnamefont
  {Br{\"a}uhaus}}, \bibinfo {author} {\bibfnamefont {U.}~\bibnamefont
  {Schr{\"o}der}}, \ and\ \bibinfo {author} {\bibfnamefont {U.}~\bibnamefont
  {B{\"o}ttger}},\ }\href@noop {} {\bibfield  {journal} {\bibinfo  {journal}
  {Appl. Phys. Lett.}\ }\textbf {\bibinfo {volume} {99}},\ \bibinfo {pages}
  {102903} (\bibinfo {year} {2011}{\natexlab{a}})}\BibitemShut {NoStop}%
\bibitem [{\citenamefont {B{\"o}scke}\ \emph
  {et~al.}(2011{\natexlab{b}})\citenamefont {B{\"o}scke}, \citenamefont
  {Teichert}, \citenamefont {Br{\"a}uhaus}, \citenamefont {M{\"u}ller},
  \citenamefont {Schr{\"o}der}, \citenamefont {B{\"o}ttger},\ and\
  \citenamefont {Mikolajick}}]{Boscke11p112904}%
  \BibitemOpen
  \bibfield  {author} {\bibinfo {author} {\bibfnamefont {T.~S.}\ \bibnamefont
  {B{\"o}scke}}, \bibinfo {author} {\bibfnamefont {S.}~\bibnamefont
  {Teichert}}, \bibinfo {author} {\bibfnamefont {D.}~\bibnamefont
  {Br{\"a}uhaus}}, \bibinfo {author} {\bibfnamefont {J.}~\bibnamefont
  {M{\"u}ller}}, \bibinfo {author} {\bibfnamefont {U.}~\bibnamefont
  {Schr{\"o}der}}, \bibinfo {author} {\bibfnamefont {U.}~\bibnamefont
  {B{\"o}ttger}}, \ and\ \bibinfo {author} {\bibfnamefont {T.}~\bibnamefont
  {Mikolajick}},\ }\href@noop {} {\bibfield  {journal} {\bibinfo  {journal}
  {Appl. Phys. Lett.}\ }\textbf {\bibinfo {volume} {99}},\ \bibinfo {pages}
  {112904} (\bibinfo {year} {2011}{\natexlab{b}})}\BibitemShut {NoStop}%
\bibitem [{\citenamefont {Mueller}\ \emph
  {et~al.}(2012{\natexlab{a}})\citenamefont {Mueller}, \citenamefont
  {Adelmann}, \citenamefont {Singh}, \citenamefont {Van~Elshocht},
  \citenamefont {Schroeder},\ and\ \citenamefont {Mikolajick}}]{Mueller12p123}%
  \BibitemOpen
  \bibfield  {author} {\bibinfo {author} {\bibfnamefont {S.}~\bibnamefont
  {Mueller}}, \bibinfo {author} {\bibfnamefont {C.}~\bibnamefont {Adelmann}},
  \bibinfo {author} {\bibfnamefont {A.}~\bibnamefont {Singh}}, \bibinfo
  {author} {\bibfnamefont {S.}~\bibnamefont {Van~Elshocht}}, \bibinfo {author}
  {\bibfnamefont {U.}~\bibnamefont {Schroeder}}, \ and\ \bibinfo {author}
  {\bibfnamefont {T.}~\bibnamefont {Mikolajick}},\ }\href@noop {} {\bibfield
  {journal} {\bibinfo  {journal} {ECS J. Solid State Sci. Technol.}\ }\textbf
  {\bibinfo {volume} {1}},\ \bibinfo {pages} {N123} (\bibinfo {year}
  {2012}{\natexlab{a}})}\BibitemShut {NoStop}%
\bibitem [{\citenamefont {Mueller}\ \emph
  {et~al.}(2012{\natexlab{b}})\citenamefont {Mueller}, \citenamefont {Mueller},
  \citenamefont {Singh}, \citenamefont {Riedel}, \citenamefont {Sundqvist},
  \citenamefont {Schroeder},\ and\ \citenamefont
  {Mikolajick}}]{Mueller12p2412}%
  \BibitemOpen
  \bibfield  {author} {\bibinfo {author} {\bibfnamefont {S.}~\bibnamefont
  {Mueller}}, \bibinfo {author} {\bibfnamefont {J.}~\bibnamefont {Mueller}},
  \bibinfo {author} {\bibfnamefont {A.}~\bibnamefont {Singh}}, \bibinfo
  {author} {\bibfnamefont {S.}~\bibnamefont {Riedel}}, \bibinfo {author}
  {\bibfnamefont {J.}~\bibnamefont {Sundqvist}}, \bibinfo {author}
  {\bibfnamefont {U.}~\bibnamefont {Schroeder}}, \ and\ \bibinfo {author}
  {\bibfnamefont {T.}~\bibnamefont {Mikolajick}},\ }\href@noop {} {\bibfield
  {journal} {\bibinfo  {journal} {Adv. Funct. Mater.}\ }\textbf {\bibinfo
  {volume} {22}},\ \bibinfo {pages} {2412} (\bibinfo {year}
  {2012}{\natexlab{b}})}\BibitemShut {NoStop}%
\bibitem [{\citenamefont {M{\"u}ller}\ \emph
  {et~al.}(2011{\natexlab{a}})\citenamefont {M{\"u}ller}, \citenamefont
  {B{\"o}scke}, \citenamefont {Br{\"a}uhaus}, \citenamefont {Schr{\"o}der},
  \citenamefont {B{\"o}ttger}, \citenamefont {Sundqvist}, \citenamefont
  {K{\"u}cher}, \citenamefont {Mikolajick},\ and\ \citenamefont
  {Frey}}]{Muller11p112901}%
  \BibitemOpen
  \bibfield  {author} {\bibinfo {author} {\bibfnamefont {J.}~\bibnamefont
  {M{\"u}ller}}, \bibinfo {author} {\bibfnamefont {T.}~\bibnamefont
  {B{\"o}scke}}, \bibinfo {author} {\bibfnamefont {D.}~\bibnamefont
  {Br{\"a}uhaus}}, \bibinfo {author} {\bibfnamefont {U.}~\bibnamefont
  {Schr{\"o}der}}, \bibinfo {author} {\bibfnamefont {U.}~\bibnamefont
  {B{\"o}ttger}}, \bibinfo {author} {\bibfnamefont {J.}~\bibnamefont
  {Sundqvist}}, \bibinfo {author} {\bibfnamefont {P.}~\bibnamefont
  {K{\"u}cher}}, \bibinfo {author} {\bibfnamefont {T.}~\bibnamefont
  {Mikolajick}}, \ and\ \bibinfo {author} {\bibfnamefont {L.}~\bibnamefont
  {Frey}},\ }\href@noop {} {\bibfield  {journal} {\bibinfo  {journal} {Appl.
  Phys. Lett.}\ }\textbf {\bibinfo {volume} {99}},\ \bibinfo {pages} {112901}
  (\bibinfo {year} {2011}{\natexlab{a}})}\BibitemShut {NoStop}%
\bibitem [{\citenamefont {M{\"u}ller}\ \emph
  {et~al.}(2011{\natexlab{b}})\citenamefont {M{\"u}ller}, \citenamefont
  {Schr{\"o}der}, \citenamefont {B{\"o}scke}, \citenamefont {M{\"u}ller},
  \citenamefont {B{\"o}ttger}, \citenamefont {Wilde}, \citenamefont
  {Sundqvist}, \citenamefont {Lemberger}, \citenamefont {K{\"u}cher},
  \citenamefont {Mikolajick} \emph {et~al.}}]{Muller11p114113}%
  \BibitemOpen
  \bibfield  {author} {\bibinfo {author} {\bibfnamefont {J.}~\bibnamefont
  {M{\"u}ller}}, \bibinfo {author} {\bibfnamefont {U.}~\bibnamefont
  {Schr{\"o}der}}, \bibinfo {author} {\bibfnamefont {T.}~\bibnamefont
  {B{\"o}scke}}, \bibinfo {author} {\bibfnamefont {I.}~\bibnamefont
  {M{\"u}ller}}, \bibinfo {author} {\bibfnamefont {U.}~\bibnamefont
  {B{\"o}ttger}}, \bibinfo {author} {\bibfnamefont {L.}~\bibnamefont {Wilde}},
  \bibinfo {author} {\bibfnamefont {J.}~\bibnamefont {Sundqvist}}, \bibinfo
  {author} {\bibfnamefont {M.}~\bibnamefont {Lemberger}}, \bibinfo {author}
  {\bibfnamefont {P.}~\bibnamefont {K{\"u}cher}}, \bibinfo {author}
  {\bibfnamefont {T.}~\bibnamefont {Mikolajick}},  \emph {et~al.},\ }\href@noop
  {} {\bibfield  {journal} {\bibinfo  {journal} {J. Appl. Phys.}\ }\textbf
  {\bibinfo {volume} {110}},\ \bibinfo {pages} {114113} (\bibinfo {year}
  {2011}{\natexlab{b}})}\BibitemShut {NoStop}%
\bibitem [{\citenamefont {Müller}\ \emph {et~al.}(2012)\citenamefont
  {Müller}, \citenamefont {Böscke}, \citenamefont {Schröder},
  \citenamefont {Mueller}, \citenamefont {Bräuhaus}, \citenamefont
  {Böttger}, \citenamefont {Frey},\ and\ \citenamefont
  {Mikolajick}}]{Muller12p4318}%
  \BibitemOpen
  \bibfield  {author} {\bibinfo {author} {\bibfnamefont {J.}~\bibnamefont
  {Müller}}, \bibinfo {author} {\bibfnamefont {T.~S.}\ \bibnamefont
  {Böscke}}, \bibinfo {author} {\bibfnamefont {U.}~\bibnamefont
  {Schröder}}, \bibinfo {author} {\bibfnamefont {S.}~\bibnamefont {Mueller}},
  \bibinfo {author} {\bibfnamefont {D.}~\bibnamefont {Bräuhaus}}, \bibinfo
  {author} {\bibfnamefont {U.}~\bibnamefont {Böttger}}, \bibinfo {author}
  {\bibfnamefont {L.}~\bibnamefont {Frey}}, \ and\ \bibinfo {author}
  {\bibfnamefont {T.}~\bibnamefont {Mikolajick}},\ }\href@noop {} {\bibfield
  {journal} {\bibinfo  {journal} {Nano Lett.}\ }\textbf {\bibinfo {volume}
  {12}},\ \bibinfo {pages} {4318} (\bibinfo {year} {2012})}\BibitemShut
  {NoStop}%
\bibitem [{\citenamefont {Ruh}\ \emph {et~al.}(1968)\citenamefont {Ruh},
  \citenamefont {Garrett}, \citenamefont {Domagala},\ and\ \citenamefont
  {Tallan}}]{Ruh68p23}%
  \BibitemOpen
  \bibfield  {author} {\bibinfo {author} {\bibfnamefont {R.}~\bibnamefont
  {Ruh}}, \bibinfo {author} {\bibfnamefont {H.~J.}\ \bibnamefont {Garrett}},
  \bibinfo {author} {\bibfnamefont {R.~F.}\ \bibnamefont {Domagala}}, \ and\
  \bibinfo {author} {\bibfnamefont {N.~M.}\ \bibnamefont {Tallan}},\
  }\href@noop {} {\bibfield  {journal} {\bibinfo  {journal} {J. Am. Ceram.
  Soc.}\ }\textbf {\bibinfo {volume} {51}},\ \bibinfo {pages} {23} (\bibinfo
  {year} {1968})}\BibitemShut {NoStop}%
\bibitem [{\citenamefont {Reyes-Lillo}\ \emph {et~al.}(2014)\citenamefont
  {Reyes-Lillo}, \citenamefont {Garrity},\ and\ \citenamefont
  {Rabe}}]{Reyes14p140103}%
  \BibitemOpen
  \bibfield  {author} {\bibinfo {author} {\bibfnamefont {S.~E.}\ \bibnamefont
  {Reyes-Lillo}}, \bibinfo {author} {\bibfnamefont {K.~F.}\ \bibnamefont
  {Garrity}}, \ and\ \bibinfo {author} {\bibfnamefont {K.~M.}\ \bibnamefont
  {Rabe}},\ }\href@noop {} {\bibfield  {journal} {\bibinfo  {journal} {Phys.
  Rev. B}\ }\textbf {\bibinfo {volume} {90}},\ \bibinfo {pages} {140103}
  (\bibinfo {year} {2014})}\BibitemShut {NoStop}%
\bibitem [{\citenamefont {Lee}\ \emph {et~al.}(2020)\citenamefont {Lee},
  \citenamefont {Lee}, \citenamefont {Lee}, \citenamefont {Jo}, \citenamefont
  {Yang}, \citenamefont {Kim}, \citenamefont {Chae}, \citenamefont {Waghmare},\
  and\ \citenamefont {Lee}}]{Lee20p1}%
  \BibitemOpen
  \bibfield  {author} {\bibinfo {author} {\bibfnamefont {H.-J.}\ \bibnamefont
  {Lee}}, \bibinfo {author} {\bibfnamefont {M.}~\bibnamefont {Lee}}, \bibinfo
  {author} {\bibfnamefont {K.}~\bibnamefont {Lee}}, \bibinfo {author}
  {\bibfnamefont {J.}~\bibnamefont {Jo}}, \bibinfo {author} {\bibfnamefont
  {H.}~\bibnamefont {Yang}}, \bibinfo {author} {\bibfnamefont {Y.}~\bibnamefont
  {Kim}}, \bibinfo {author} {\bibfnamefont {S.~C.}\ \bibnamefont {Chae}},
  \bibinfo {author} {\bibfnamefont {U.}~\bibnamefont {Waghmare}}, \ and\
  \bibinfo {author} {\bibfnamefont {J.~H.}\ \bibnamefont {Lee}},\ }\href@noop
  {} {\bibfield  {journal} {\bibinfo  {journal} {Science}\ } (\bibinfo {year}
  {2020})}\BibitemShut {NoStop}%
\bibitem [{\citenamefont {Sang}\ \emph {et~al.}(2015)\citenamefont {Sang},
  \citenamefont {Grimley}, \citenamefont {Schenk}, \citenamefont {Schroeder},\
  and\ \citenamefont {LeBeau}}]{Sang15p162905}%
  \BibitemOpen
  \bibfield  {author} {\bibinfo {author} {\bibfnamefont {X.}~\bibnamefont
  {Sang}}, \bibinfo {author} {\bibfnamefont {E.~D.}\ \bibnamefont {Grimley}},
  \bibinfo {author} {\bibfnamefont {T.}~\bibnamefont {Schenk}}, \bibinfo
  {author} {\bibfnamefont {U.}~\bibnamefont {Schroeder}}, \ and\ \bibinfo
  {author} {\bibfnamefont {J.~M.}\ \bibnamefont {LeBeau}},\ }\href@noop {}
  {\bibfield  {journal} {\bibinfo  {journal} {Appl. Phys. Lett.}\ }\textbf
  {\bibinfo {volume} {106}},\ \bibinfo {pages} {162905} (\bibinfo {year}
  {2015})}\BibitemShut {NoStop}%
\bibitem [{\citenamefont {Park}\ \emph
  {et~al.}(2015{\natexlab{a}})\citenamefont {Park}, \citenamefont {Lee},
  \citenamefont {Kim}, \citenamefont {Kim}, \citenamefont {Moon}, \citenamefont
  {Kim}, \citenamefont {M{\"u}ller}, \citenamefont {Kersch}, \citenamefont
  {Schroeder}, \citenamefont {Mikolajick} \emph {et~al.}}]{Park15p1811}%
  \BibitemOpen
  \bibfield  {author} {\bibinfo {author} {\bibfnamefont {M.~H.}\ \bibnamefont
  {Park}}, \bibinfo {author} {\bibfnamefont {Y.~H.}\ \bibnamefont {Lee}},
  \bibinfo {author} {\bibfnamefont {H.~J.}\ \bibnamefont {Kim}}, \bibinfo
  {author} {\bibfnamefont {Y.~J.}\ \bibnamefont {Kim}}, \bibinfo {author}
  {\bibfnamefont {T.}~\bibnamefont {Moon}}, \bibinfo {author} {\bibfnamefont
  {K.~D.}\ \bibnamefont {Kim}}, \bibinfo {author} {\bibfnamefont
  {J.}~\bibnamefont {M{\"u}ller}}, \bibinfo {author} {\bibfnamefont
  {A.}~\bibnamefont {Kersch}}, \bibinfo {author} {\bibfnamefont
  {U.}~\bibnamefont {Schroeder}}, \bibinfo {author} {\bibfnamefont
  {T.}~\bibnamefont {Mikolajick}},  \emph {et~al.},\ }\href@noop {} {\bibfield
  {journal} {\bibinfo  {journal} {Adv. Mater.}\ }\textbf {\bibinfo {volume}
  {27}},\ \bibinfo {pages} {1811} (\bibinfo {year}
  {2015}{\natexlab{a}})}\BibitemShut {NoStop}%
\bibitem [{\citenamefont {Huan}\ \emph {et~al.}(2014)\citenamefont {Huan},
  \citenamefont {Sharma}, \citenamefont {Rossetti~Jr},\ and\ \citenamefont
  {Ramprasad}}]{Huan14p064111}%
  \BibitemOpen
  \bibfield  {author} {\bibinfo {author} {\bibfnamefont {T.~D.}\ \bibnamefont
  {Huan}}, \bibinfo {author} {\bibfnamefont {V.}~\bibnamefont {Sharma}},
  \bibinfo {author} {\bibfnamefont {G.~A.}\ \bibnamefont {Rossetti~Jr}}, \ and\
  \bibinfo {author} {\bibfnamefont {R.}~\bibnamefont {Ramprasad}},\ }\href@noop
  {} {\bibfield  {journal} {\bibinfo  {journal} {Phys. Rev. B}\ }\textbf
  {\bibinfo {volume} {90}},\ \bibinfo {pages} {064111} (\bibinfo {year}
  {2014})}\BibitemShut {NoStop}%
\bibitem [{\citenamefont {Giannozzi}\ \emph {et~al.}(2009)\citenamefont
  {Giannozzi}, \citenamefont {Baroni}, \citenamefont {Bonini}, \citenamefont
  {Calandra}, \citenamefont {Car}, \citenamefont {Cavazzoni}, \citenamefont
  {Ceresoli}, \citenamefont {Chiarotti}, \citenamefont {Cococcioni},
  \citenamefont {Dabo}, \citenamefont {Corso}, \citenamefont {de~Gironcoli},
  \citenamefont {Fabris}, \citenamefont {Fratesi}, \citenamefont {Gebauer},
  \citenamefont {Gerstmann}, \citenamefont {Gougoussis}, \citenamefont
  {Kokalj}, \citenamefont {Lazzeri}, \citenamefont {Martin-Samos},
  \citenamefont {Marzari}, \citenamefont {Mauri}, \citenamefont {Mazzarello},
  \citenamefont {Paolini}, \citenamefont {Pasquarello}, \citenamefont
  {Paulatto}, \citenamefont {Sbraccia}, \citenamefont {Scandolo}, \citenamefont
  {Sclauzero}, \citenamefont {Seitsonen}, \citenamefont {Smogunov},
  \citenamefont {Umari},\ and\ \citenamefont
  {Wentzcovitch}}]{Giannozzi09p395502etalp}%
  \BibitemOpen
  \bibfield  {author} {\bibinfo {author} {\bibfnamefont {P.}~\bibnamefont
  {Giannozzi}}, \bibinfo {author} {\bibfnamefont {S.}~\bibnamefont {Baroni}},
  \bibinfo {author} {\bibfnamefont {N.}~\bibnamefont {Bonini}}, \bibinfo
  {author} {\bibfnamefont {M.}~\bibnamefont {Calandra}}, \bibinfo {author}
  {\bibfnamefont {R.}~\bibnamefont {Car}}, \bibinfo {author} {\bibfnamefont
  {C.}~\bibnamefont {Cavazzoni}}, \bibinfo {author} {\bibfnamefont
  {D.}~\bibnamefont {Ceresoli}}, \bibinfo {author} {\bibfnamefont {G.~L.}\
  \bibnamefont {Chiarotti}}, \bibinfo {author} {\bibfnamefont {M.}~\bibnamefont
  {Cococcioni}}, \bibinfo {author} {\bibfnamefont {I.}~\bibnamefont {Dabo}},
  \bibinfo {author} {\bibfnamefont {A.~D.}\ \bibnamefont {Corso}}, \bibinfo
  {author} {\bibfnamefont {S.}~\bibnamefont {de~Gironcoli}}, \bibinfo {author}
  {\bibfnamefont {S.}~\bibnamefont {Fabris}}, \bibinfo {author} {\bibfnamefont
  {G.}~\bibnamefont {Fratesi}}, \bibinfo {author} {\bibfnamefont
  {R.}~\bibnamefont {Gebauer}}, \bibinfo {author} {\bibfnamefont
  {U.}~\bibnamefont {Gerstmann}}, \bibinfo {author} {\bibfnamefont
  {C.}~\bibnamefont {Gougoussis}}, \bibinfo {author} {\bibfnamefont
  {A.}~\bibnamefont {Kokalj}}, \bibinfo {author} {\bibfnamefont
  {M.}~\bibnamefont {Lazzeri}}, \bibinfo {author} {\bibfnamefont
  {L.}~\bibnamefont {Martin-Samos}}, \bibinfo {author} {\bibfnamefont
  {N.}~\bibnamefont {Marzari}}, \bibinfo {author} {\bibfnamefont
  {F.}~\bibnamefont {Mauri}}, \bibinfo {author} {\bibfnamefont
  {R.}~\bibnamefont {Mazzarello}}, \bibinfo {author} {\bibfnamefont
  {S.}~\bibnamefont {Paolini}}, \bibinfo {author} {\bibfnamefont
  {A.}~\bibnamefont {Pasquarello}}, \bibinfo {author} {\bibfnamefont
  {L.}~\bibnamefont {Paulatto}}, \bibinfo {author} {\bibfnamefont
  {C.}~\bibnamefont {Sbraccia}}, \bibinfo {author} {\bibfnamefont
  {S.}~\bibnamefont {Scandolo}}, \bibinfo {author} {\bibfnamefont
  {G.}~\bibnamefont {Sclauzero}}, \bibinfo {author} {\bibfnamefont {A.~P.}\
  \bibnamefont {Seitsonen}}, \bibinfo {author} {\bibfnamefont {A.}~\bibnamefont
  {Smogunov}}, \bibinfo {author} {\bibfnamefont {P.}~\bibnamefont {Umari}}, \
  and\ \bibinfo {author} {\bibfnamefont {R.~M.}\ \bibnamefont {Wentzcovitch}},\
  }\href@noop {} {\bibfield  {journal} {\bibinfo  {journal} {J. Phys.: Condens.
  Matter}\ }\textbf {\bibinfo {volume} {21}},\ \bibinfo {pages} {395502}
  (\bibinfo {year} {2009})}\BibitemShut {NoStop}%
\bibitem [{\citenamefont {Gonze}\ \emph {et~al.}(2002)\citenamefont {Gonze},
  \citenamefont {Beuken}, \citenamefont {Caracas}, \citenamefont {Detraux},
  \citenamefont {Fuchs}, \citenamefont {Rignanese}, \citenamefont {Sindic},
  \citenamefont {Verstraete}, \citenamefont {Zerah}, \citenamefont {Jollet},
  \citenamefont {Torrent}, \citenamefont {Roy}, \citenamefont {Mikami},
  \citenamefont {Ghosez.}, \citenamefont {Raty},\ and\ \citenamefont
  {Allan}}]{Gonze02p478}%
  \BibitemOpen
  \bibfield  {author} {\bibinfo {author} {\bibfnamefont {X.}~\bibnamefont
  {Gonze}}, \bibinfo {author} {\bibfnamefont {J.-M.}\ \bibnamefont {Beuken}},
  \bibinfo {author} {\bibfnamefont {R.}~\bibnamefont {Caracas}}, \bibinfo
  {author} {\bibfnamefont {F.}~\bibnamefont {Detraux}}, \bibinfo {author}
  {\bibfnamefont {M.}~\bibnamefont {Fuchs}}, \bibinfo {author} {\bibfnamefont
  {G.-M.}\ \bibnamefont {Rignanese}}, \bibinfo {author} {\bibfnamefont
  {L.}~\bibnamefont {Sindic}}, \bibinfo {author} {\bibfnamefont
  {M.}~\bibnamefont {Verstraete}}, \bibinfo {author} {\bibfnamefont
  {G.}~\bibnamefont {Zerah}}, \bibinfo {author} {\bibfnamefont
  {F.}~\bibnamefont {Jollet}}, \bibinfo {author} {\bibfnamefont
  {M.}~\bibnamefont {Torrent}}, \bibinfo {author} {\bibfnamefont
  {A.}~\bibnamefont {Roy}}, \bibinfo {author} {\bibfnamefont {M.}~\bibnamefont
  {Mikami}}, \bibinfo {author} {\bibfnamefont {P.}~\bibnamefont {Ghosez.}},
  \bibinfo {author} {\bibfnamefont {J.-Y.}\ \bibnamefont {Raty}}, \ and\
  \bibinfo {author} {\bibfnamefont {D.}~\bibnamefont {Allan}},\ }\href@noop {}
  {\bibfield  {journal} {\bibinfo  {journal} {Comp. Mater. Sci.}\ }\textbf
  {\bibinfo {volume} {25}},\ \bibinfo {pages} {478} (\bibinfo {year}
  {2002})}\BibitemShut {NoStop}%
\bibitem [{Opi()}]{Opium}%
  \BibitemOpen
  \href@noop {} {}\bibinfo {howpublished}
  {http://opium.sourceforge.net}\BibitemShut {NoStop}%
\bibitem [{\citenamefont {Bennett}(2012)}]{Bennett12p14}%
  \BibitemOpen
  \bibfield  {author} {\bibinfo {author} {\bibfnamefont {J.~W.}\ \bibnamefont
  {Bennett}},\ }\href@noop {} {\bibfield  {journal} {\bibinfo  {journal} {Phys.
  Procedia}\ }\textbf {\bibinfo {volume} {34}},\ \bibinfo {pages} {14}
  (\bibinfo {year} {2012})}\BibitemShut {NoStop}%
\bibitem [{\citenamefont {Monkhorst}\ and\ \citenamefont
  {Pack}(1976)}]{Monkhorst76p5188}%
  \BibitemOpen
  \bibfield  {author} {\bibinfo {author} {\bibfnamefont {H.~J.}\ \bibnamefont
  {Monkhorst}}\ and\ \bibinfo {author} {\bibfnamefont {J.~D.}\ \bibnamefont
  {Pack}},\ }\href@noop {} {\bibfield  {journal} {\bibinfo  {journal} {Phys.
  Rev. B}\ }\textbf {\bibinfo {volume} {13}},\ \bibinfo {pages} {5188}
  (\bibinfo {year} {1976})}\BibitemShut {NoStop}%
\bibitem [{\citenamefont {Weeks}\ \emph {et~al.}(2017)\citenamefont {Weeks},
  \citenamefont {Pal}, \citenamefont {Narasimhan}, \citenamefont {Littau},\
  and\ \citenamefont {Chiang}}]{Weeks17p13440}%
  \BibitemOpen
  \bibfield  {author} {\bibinfo {author} {\bibfnamefont {S.~L.}\ \bibnamefont
  {Weeks}}, \bibinfo {author} {\bibfnamefont {A.}~\bibnamefont {Pal}}, \bibinfo
  {author} {\bibfnamefont {V.~K.}\ \bibnamefont {Narasimhan}}, \bibinfo
  {author} {\bibfnamefont {K.~A.}\ \bibnamefont {Littau}}, \ and\ \bibinfo
  {author} {\bibfnamefont {T.}~\bibnamefont {Chiang}},\ }\href@noop {}
  {\bibfield  {journal} {\bibinfo  {journal} {ACS Appl. Mater. Interfaces}\
  }\textbf {\bibinfo {volume} {9}},\ \bibinfo {pages} {13440} (\bibinfo {year}
  {2017})}\BibitemShut {NoStop}%
\bibitem [{\citenamefont {Rabe}\ \emph {et~al.}(2007)\citenamefont {Rabe},
  \citenamefont {Ahn},\ and\ \citenamefont {Triscone}}]{Rabe07}%
  \BibitemOpen
  \bibfield  {author} {\bibinfo {author} {\bibfnamefont {K.~M.}\ \bibnamefont
  {Rabe}}, \bibinfo {author} {\bibfnamefont {C.~H.}\ \bibnamefont {Ahn}}, \
  and\ \bibinfo {author} {\bibfnamefont {J.-M.}\ \bibnamefont {Triscone}},\
  }\href@noop {} {\emph {\bibinfo {title} {Physics of Ferroelectrics: A Modern
  Perspective}}}\ (\bibinfo  {publisher} {Springer--Verlag},\ \bibinfo {year}
  {2007})\BibitemShut {NoStop}%
\bibitem [{\citenamefont {Sai}\ \emph {et~al.}(2002)\citenamefont {Sai},
  \citenamefont {Rabe},\ and\ \citenamefont {Vanderbilt}}]{Sai02p104108}%
  \BibitemOpen
  \bibfield  {author} {\bibinfo {author} {\bibfnamefont {N.}~\bibnamefont
  {Sai}}, \bibinfo {author} {\bibfnamefont {K.~M.}\ \bibnamefont {Rabe}}, \
  and\ \bibinfo {author} {\bibfnamefont {D.}~\bibnamefont {Vanderbilt}},\
  }\href@noop {} {\bibfield  {journal} {\bibinfo  {journal} {Phys. Rev. B}\
  }\textbf {\bibinfo {volume} {66}},\ \bibinfo {pages} {104108} (\bibinfo
  {year} {2002})}\BibitemShut {NoStop}%
\bibitem [{\citenamefont {Lee}\ \emph {et~al.}(2008)\citenamefont {Lee},
  \citenamefont {Cho}, \citenamefont {Lee}, \citenamefont {Hwang},\ and\
  \citenamefont {Han}}]{Lee08p012102}%
  \BibitemOpen
  \bibfield  {author} {\bibinfo {author} {\bibfnamefont {C.-K.}\ \bibnamefont
  {Lee}}, \bibinfo {author} {\bibfnamefont {E.}~\bibnamefont {Cho}}, \bibinfo
  {author} {\bibfnamefont {H.-S.}\ \bibnamefont {Lee}}, \bibinfo {author}
  {\bibfnamefont {C.~S.}\ \bibnamefont {Hwang}}, \ and\ \bibinfo {author}
  {\bibfnamefont {S.}~\bibnamefont {Han}},\ }\href@noop {} {\bibfield
  {journal} {\bibinfo  {journal} {Phys. Rev. B}\ }\textbf {\bibinfo {volume}
  {78}},\ \bibinfo {pages} {012102} (\bibinfo {year} {2008})}\BibitemShut
  {NoStop}%
\bibitem [{\citenamefont {Zhou}\ \emph {et~al.}(2013)\citenamefont {Zhou},
  \citenamefont {Xu}, \citenamefont {Li}, \citenamefont {Guan}, \citenamefont
  {Cao}, \citenamefont {Dong}, \citenamefont {M{\"u}ller}, \citenamefont
  {Schenk},\ and\ \citenamefont {Schr{\"o}der}}]{Zhou2013p192904}%
  \BibitemOpen
  \bibfield  {author} {\bibinfo {author} {\bibfnamefont {D.}~\bibnamefont
  {Zhou}}, \bibinfo {author} {\bibfnamefont {J.}~\bibnamefont {Xu}}, \bibinfo
  {author} {\bibfnamefont {Q.}~\bibnamefont {Li}}, \bibinfo {author}
  {\bibfnamefont {Y.}~\bibnamefont {Guan}}, \bibinfo {author} {\bibfnamefont
  {F.}~\bibnamefont {Cao}}, \bibinfo {author} {\bibfnamefont {X.}~\bibnamefont
  {Dong}}, \bibinfo {author} {\bibfnamefont {J.}~\bibnamefont {M{\"u}ller}},
  \bibinfo {author} {\bibfnamefont {T.}~\bibnamefont {Schenk}}, \ and\ \bibinfo
  {author} {\bibfnamefont {U.}~\bibnamefont {Schr{\"o}der}},\ }\href@noop {}
  {\bibfield  {journal} {\bibinfo  {journal} {Appl. Phys. Lett.}\ }\textbf
  {\bibinfo {volume} {103}},\ \bibinfo {pages} {192904} (\bibinfo {year}
  {2013})}\BibitemShut {NoStop}%
\bibitem [{\citenamefont {Pe{\v{s}}i{\'c}}\ \emph {et~al.}(2016)\citenamefont
  {Pe{\v{s}}i{\'c}}, \citenamefont {Fengler}, \citenamefont {Larcher},
  \citenamefont {Padovani}, \citenamefont {Schenk}, \citenamefont {Grimley},
  \citenamefont {Sang}, \citenamefont {LeBeau}, \citenamefont {Slesazeck},
  \citenamefont {Schroeder} \emph {et~al.}}]{Pevsic16p4601}%
  \BibitemOpen
  \bibfield  {author} {\bibinfo {author} {\bibfnamefont {M.}~\bibnamefont
  {Pe{\v{s}}i{\'c}}}, \bibinfo {author} {\bibfnamefont {F.~P.~G.}\ \bibnamefont
  {Fengler}}, \bibinfo {author} {\bibfnamefont {L.}~\bibnamefont {Larcher}},
  \bibinfo {author} {\bibfnamefont {A.}~\bibnamefont {Padovani}}, \bibinfo
  {author} {\bibfnamefont {T.}~\bibnamefont {Schenk}}, \bibinfo {author}
  {\bibfnamefont {E.~D.}\ \bibnamefont {Grimley}}, \bibinfo {author}
  {\bibfnamefont {X.}~\bibnamefont {Sang}}, \bibinfo {author} {\bibfnamefont
  {J.~M.}\ \bibnamefont {LeBeau}}, \bibinfo {author} {\bibfnamefont
  {S.}~\bibnamefont {Slesazeck}}, \bibinfo {author} {\bibfnamefont
  {U.}~\bibnamefont {Schroeder}},  \emph {et~al.},\ }\href@noop {} {\bibfield
  {journal} {\bibinfo  {journal} {Adv. Funct. Mater.}\ }\textbf {\bibinfo
  {volume} {26}},\ \bibinfo {pages} {4601} (\bibinfo {year}
  {2016})}\BibitemShut {NoStop}%
\bibitem [{\citenamefont {Grimley}\ \emph {et~al.}(2016)\citenamefont
  {Grimley}, \citenamefont {Schenk}, \citenamefont {Sang}, \citenamefont
  {Pe{\v{s}}i{\'c}}, \citenamefont {Schroeder}, \citenamefont {Mikolajick},\
  and\ \citenamefont {LeBeau}}]{Grimley16p1600173}%
  \BibitemOpen
  \bibfield  {author} {\bibinfo {author} {\bibfnamefont {E.~D.}\ \bibnamefont
  {Grimley}}, \bibinfo {author} {\bibfnamefont {T.}~\bibnamefont {Schenk}},
  \bibinfo {author} {\bibfnamefont {X.}~\bibnamefont {Sang}}, \bibinfo {author}
  {\bibfnamefont {M.}~\bibnamefont {Pe{\v{s}}i{\'c}}}, \bibinfo {author}
  {\bibfnamefont {U.}~\bibnamefont {Schroeder}}, \bibinfo {author}
  {\bibfnamefont {T.}~\bibnamefont {Mikolajick}}, \ and\ \bibinfo {author}
  {\bibfnamefont {J.~M.}\ \bibnamefont {LeBeau}},\ }\href@noop {} {\bibfield
  {journal} {\bibinfo  {journal} {Adv. Electron. Mater.}\ }\textbf {\bibinfo
  {volume} {2}},\ \bibinfo {pages} {1600173} (\bibinfo {year}
  {2016})}\BibitemShut {NoStop}%
\bibitem [{\citenamefont {Martin}\ \emph {et~al.}(2014)\citenamefont {Martin},
  \citenamefont {M{\"u}ller}, \citenamefont {Schenk}, \citenamefont {Arruda},
  \citenamefont {Kumar}, \citenamefont {Strelcov}, \citenamefont {Yurchuk},
  \citenamefont {M{\"u}ller}, \citenamefont {Pohl}, \citenamefont
  {Schr{\"o}der} \emph {et~al.}}]{Martin14p8198}%
  \BibitemOpen
  \bibfield  {author} {\bibinfo {author} {\bibfnamefont {D.}~\bibnamefont
  {Martin}}, \bibinfo {author} {\bibfnamefont {J.}~\bibnamefont {M{\"u}ller}},
  \bibinfo {author} {\bibfnamefont {T.}~\bibnamefont {Schenk}}, \bibinfo
  {author} {\bibfnamefont {T.~M.}\ \bibnamefont {Arruda}}, \bibinfo {author}
  {\bibfnamefont {A.}~\bibnamefont {Kumar}}, \bibinfo {author} {\bibfnamefont
  {E.}~\bibnamefont {Strelcov}}, \bibinfo {author} {\bibfnamefont
  {E.}~\bibnamefont {Yurchuk}}, \bibinfo {author} {\bibfnamefont
  {S.}~\bibnamefont {M{\"u}ller}}, \bibinfo {author} {\bibfnamefont
  {D.}~\bibnamefont {Pohl}}, \bibinfo {author} {\bibfnamefont {U.}~\bibnamefont
  {Schr{\"o}der}},  \emph {et~al.},\ }\href@noop {} {\bibfield  {journal}
  {\bibinfo  {journal} {Adv. Mater.}\ }\textbf {\bibinfo {volume} {26}},\
  \bibinfo {pages} {8198} (\bibinfo {year} {2014})}\BibitemShut {NoStop}%
\bibitem [{\citenamefont {Hoffmann}\ \emph
  {et~al.}(2015{\natexlab{a}})\citenamefont {Hoffmann}, \citenamefont
  {Schroeder}, \citenamefont {Schenk}, \citenamefont {Shimizu}, \citenamefont
  {Funakubo}, \citenamefont {Sakata}, \citenamefont {Pohl}, \citenamefont
  {Drescher}, \citenamefont {Adelmann}, \citenamefont {Materlik} \emph
  {et~al.}}]{Hoffmann15p072006}%
  \BibitemOpen
  \bibfield  {author} {\bibinfo {author} {\bibfnamefont {M.}~\bibnamefont
  {Hoffmann}}, \bibinfo {author} {\bibfnamefont {U.}~\bibnamefont {Schroeder}},
  \bibinfo {author} {\bibfnamefont {T.}~\bibnamefont {Schenk}}, \bibinfo
  {author} {\bibfnamefont {T.}~\bibnamefont {Shimizu}}, \bibinfo {author}
  {\bibfnamefont {H.}~\bibnamefont {Funakubo}}, \bibinfo {author}
  {\bibfnamefont {O.}~\bibnamefont {Sakata}}, \bibinfo {author} {\bibfnamefont
  {D.}~\bibnamefont {Pohl}}, \bibinfo {author} {\bibfnamefont {M.}~\bibnamefont
  {Drescher}}, \bibinfo {author} {\bibfnamefont {C.}~\bibnamefont {Adelmann}},
  \bibinfo {author} {\bibfnamefont {R.}~\bibnamefont {Materlik}},  \emph
  {et~al.},\ }\href@noop {} {\bibfield  {journal} {\bibinfo  {journal} {J.
  Appl. Phys.}\ }\textbf {\bibinfo {volume} {118}},\ \bibinfo {pages} {072006}
  (\bibinfo {year} {2015}{\natexlab{a}})}\BibitemShut {NoStop}%
\bibitem [{\citenamefont {Schenk}\ \emph {et~al.}(2015)\citenamefont {Schenk},
  \citenamefont {Hoffmann}, \citenamefont {Ocker}, \citenamefont {Pešić},
  \citenamefont {Mikolajick},\ and\ \citenamefont
  {Schroeder}}]{Schenk15p20224}%
  \BibitemOpen
  \bibfield  {author} {\bibinfo {author} {\bibfnamefont {T.}~\bibnamefont
  {Schenk}}, \bibinfo {author} {\bibfnamefont {M.}~\bibnamefont {Hoffmann}},
  \bibinfo {author} {\bibfnamefont {J.}~\bibnamefont {Ocker}}, \bibinfo
  {author} {\bibfnamefont {M.}~\bibnamefont {Pešić}}, \bibinfo {author}
  {\bibfnamefont {T.}~\bibnamefont {Mikolajick}}, \ and\ \bibinfo {author}
  {\bibfnamefont {U.}~\bibnamefont {Schroeder}},\ }\href@noop {} {\bibfield
  {journal} {\bibinfo  {journal} {ACS Appl. Mater. Interfaces}\ }\textbf
  {\bibinfo {volume} {7}},\ \bibinfo {pages} {20224} (\bibinfo {year}
  {2015})}\BibitemShut {NoStop}%
\bibitem [{\citenamefont {Schenk}\ \emph {et~al.}(2014)\citenamefont {Schenk},
  \citenamefont {Schroeder}, \citenamefont {Pešić}, \citenamefont
  {Popovici}, \citenamefont {Pershin},\ and\ \citenamefont
  {Mikolajick}}]{Schenk14p19744}%
  \BibitemOpen
  \bibfield  {author} {\bibinfo {author} {\bibfnamefont {T.}~\bibnamefont
  {Schenk}}, \bibinfo {author} {\bibfnamefont {U.}~\bibnamefont {Schroeder}},
  \bibinfo {author} {\bibfnamefont {M.}~\bibnamefont {Pešić}}, \bibinfo
  {author} {\bibfnamefont {M.}~\bibnamefont {Popovici}}, \bibinfo {author}
  {\bibfnamefont {Y.~V.}\ \bibnamefont {Pershin}}, \ and\ \bibinfo {author}
  {\bibfnamefont {T.}~\bibnamefont {Mikolajick}},\ }\href@noop {} {\bibfield
  {journal} {\bibinfo  {journal} {ACS Appl. Mater. Interfaces}\ }\textbf
  {\bibinfo {volume} {6}},\ \bibinfo {pages} {19744} (\bibinfo {year}
  {2014})}\BibitemShut {NoStop}%
\bibitem [{\citenamefont {Mart}\ \emph {et~al.}(2019)\citenamefont {Mart},
  \citenamefont {Kühnel}, \citenamefont {Kämpfe}, \citenamefont
  {Czernohorsky}, \citenamefont {Wiatr}, \citenamefont {Kolodinski},\ and\
  \citenamefont {Weinreich}}]{Mart19p2612}%
  \BibitemOpen
  \bibfield  {author} {\bibinfo {author} {\bibfnamefont {C.}~\bibnamefont
  {Mart}}, \bibinfo {author} {\bibfnamefont {K.}~\bibnamefont {Kühnel}},
  \bibinfo {author} {\bibfnamefont {T.}~\bibnamefont {Kämpfe}}, \bibinfo
  {author} {\bibfnamefont {M.}~\bibnamefont {Czernohorsky}}, \bibinfo {author}
  {\bibfnamefont {M.}~\bibnamefont {Wiatr}}, \bibinfo {author} {\bibfnamefont
  {S.}~\bibnamefont {Kolodinski}}, \ and\ \bibinfo {author} {\bibfnamefont
  {W.}~\bibnamefont {Weinreich}},\ }\href@noop {} {\bibfield  {journal}
  {\bibinfo  {journal} {ACS Appl. Mater. Interfaces}\ }\textbf {\bibinfo
  {volume} {1}},\ \bibinfo {pages} {2612} (\bibinfo {year} {2019})}\BibitemShut
  {NoStop}%
\bibitem [{\citenamefont {Hoffmann}\ \emph
  {et~al.}(2015{\natexlab{b}})\citenamefont {Hoffmann}, \citenamefont
  {Schroeder}, \citenamefont {K{\"u}nneth}, \citenamefont {Kersch},
  \citenamefont {Starschich}, \citenamefont {B{\"o}ttger},\ and\ \citenamefont
  {Mikolajick}}]{Hoffmann15p154}%
  \BibitemOpen
  \bibfield  {author} {\bibinfo {author} {\bibfnamefont {M.}~\bibnamefont
  {Hoffmann}}, \bibinfo {author} {\bibfnamefont {U.}~\bibnamefont {Schroeder}},
  \bibinfo {author} {\bibfnamefont {C.}~\bibnamefont {K{\"u}nneth}}, \bibinfo
  {author} {\bibfnamefont {A.}~\bibnamefont {Kersch}}, \bibinfo {author}
  {\bibfnamefont {S.}~\bibnamefont {Starschich}}, \bibinfo {author}
  {\bibfnamefont {U.}~\bibnamefont {B{\"o}ttger}}, \ and\ \bibinfo {author}
  {\bibfnamefont {T.}~\bibnamefont {Mikolajick}},\ }\href@noop {} {\bibfield
  {journal} {\bibinfo  {journal} {Nano Energy}\ }\textbf {\bibinfo {volume}
  {18}},\ \bibinfo {pages} {154} (\bibinfo {year}
  {2015}{\natexlab{b}})}\BibitemShut {NoStop}%
\bibitem [{\citenamefont {Park}\ \emph
  {et~al.}(2015{\natexlab{b}})\citenamefont {Park}, \citenamefont {Kim},
  \citenamefont {Kim}, \citenamefont {Lee}, \citenamefont {Moon}, \citenamefont
  {Kim}, \citenamefont {Hyun},\ and\ \citenamefont {Hwang}}]{Park15p192907}%
  \BibitemOpen
  \bibfield  {author} {\bibinfo {author} {\bibfnamefont {M.~H.}\ \bibnamefont
  {Park}}, \bibinfo {author} {\bibfnamefont {H.~J.}\ \bibnamefont {Kim}},
  \bibinfo {author} {\bibfnamefont {Y.~J.}\ \bibnamefont {Kim}}, \bibinfo
  {author} {\bibfnamefont {Y.~H.}\ \bibnamefont {Lee}}, \bibinfo {author}
  {\bibfnamefont {T.}~\bibnamefont {Moon}}, \bibinfo {author} {\bibfnamefont
  {K.~D.}\ \bibnamefont {Kim}}, \bibinfo {author} {\bibfnamefont {S.~D.}\
  \bibnamefont {Hyun}}, \ and\ \bibinfo {author} {\bibfnamefont {C.~S.}\
  \bibnamefont {Hwang}},\ }\href@noop {} {\bibfield  {journal} {\bibinfo
  {journal} {Appl. Phys. Lett.}\ }\textbf {\bibinfo {volume} {107}},\ \bibinfo
  {pages} {192907} (\bibinfo {year} {2015}{\natexlab{b}})}\BibitemShut
  {NoStop}%
\bibitem [{\citenamefont {Batra}\ \emph {et~al.}(2017)\citenamefont {Batra},
  \citenamefont {Huan}, \citenamefont {Jones}, \citenamefont {Rossetti~Jr},\
  and\ \citenamefont {Ramprasad}}]{Batra17p4139}%
  \BibitemOpen
  \bibfield  {author} {\bibinfo {author} {\bibfnamefont {R.}~\bibnamefont
  {Batra}}, \bibinfo {author} {\bibfnamefont {T.~D.}\ \bibnamefont {Huan}},
  \bibinfo {author} {\bibfnamefont {J.~L.}\ \bibnamefont {Jones}}, \bibinfo
  {author} {\bibfnamefont {G.}~\bibnamefont {Rossetti~Jr}}, \ and\ \bibinfo
  {author} {\bibfnamefont {R.}~\bibnamefont {Ramprasad}},\ }\href@noop {}
  {\bibfield  {journal} {\bibinfo  {journal} {J. Phys. Chem. C}\ }\textbf
  {\bibinfo {volume} {121}},\ \bibinfo {pages} {4139} (\bibinfo {year}
  {2017})}\BibitemShut {NoStop}%
\bibitem [{\citenamefont {Batra}\ \emph {et~al.}(2016)\citenamefont {Batra},
  \citenamefont {Tran},\ and\ \citenamefont {Ramprasad}}]{Batra16p172902}%
  \BibitemOpen
  \bibfield  {author} {\bibinfo {author} {\bibfnamefont {R.}~\bibnamefont
  {Batra}}, \bibinfo {author} {\bibfnamefont {H.~D.}\ \bibnamefont {Tran}}, \
  and\ \bibinfo {author} {\bibfnamefont {R.}~\bibnamefont {Ramprasad}},\
  }\href@noop {} {\bibfield  {journal} {\bibinfo  {journal} {Appl. Phys.
  Lett.}\ }\textbf {\bibinfo {volume} {108}},\ \bibinfo {pages} {172902}
  (\bibinfo {year} {2016})}\BibitemShut {NoStop}%
\bibitem [{\citenamefont {Hoffmann}\ \emph {et~al.}(2016)\citenamefont
  {Hoffmann}, \citenamefont {Pe{\v{s}}i{\'c}}, \citenamefont {Chatterjee},
  \citenamefont {Khan}, \citenamefont {Salahuddin}, \citenamefont {Slesazeck},
  \citenamefont {Schroeder},\ and\ \citenamefont
  {Mikolajick}}]{Hoffmann16p8643}%
  \BibitemOpen
  \bibfield  {author} {\bibinfo {author} {\bibfnamefont {M.}~\bibnamefont
  {Hoffmann}}, \bibinfo {author} {\bibfnamefont {M.}~\bibnamefont
  {Pe{\v{s}}i{\'c}}}, \bibinfo {author} {\bibfnamefont {K.}~\bibnamefont
  {Chatterjee}}, \bibinfo {author} {\bibfnamefont {A.~I.}\ \bibnamefont
  {Khan}}, \bibinfo {author} {\bibfnamefont {S.}~\bibnamefont {Salahuddin}},
  \bibinfo {author} {\bibfnamefont {S.}~\bibnamefont {Slesazeck}}, \bibinfo
  {author} {\bibfnamefont {U.}~\bibnamefont {Schroeder}}, \ and\ \bibinfo
  {author} {\bibfnamefont {T.}~\bibnamefont {Mikolajick}},\ }\href@noop {}
  {\bibfield  {journal} {\bibinfo  {journal} {Adv. Funct. Mater.}\ }\textbf
  {\bibinfo {volume} {26}},\ \bibinfo {pages} {8643} (\bibinfo {year}
  {2016})}\BibitemShut {NoStop}%
\end{thebibliography}%

\end{document}